\documentclass[a4paper,11pt]{article}

\usepackage{jheppub} 
\usepackage{bm}
\usepackage{soul}

\usepackage[T1]{fontenc} 

\newcommand\arccoth {\mathrm{arccoth}}

\title{\boldmath Entanglement on a Sphere}

\author[a]{K. Boutivas,}
\author[a]{D. Katsinis,}
\author[a,b]{G. Pastras,}
\author[a]{N. Prountzos}
\author[a]{and N. Tetradis}

\affiliation[a]{Department of Physics, University of Athens, Zographou 157 84, Greece}
\affiliation[b]{Laboratory for Manufacturing Systems and Automation, Department of Mechanical Engineering and Aeronautics, University of Patras, Patra 26110, Greece}

\emailAdd{kboutivas@phys.uoa.gr}
\emailAdd{dkatsinis@phys.uoa.gr}
\emailAdd{nproun@phys.uoa.gr}
\emailAdd{pastras@lms.mech.upatras.gr}
\emailAdd{ntetrad@phys.uoa.gr}

\abstract{
We study the entanglement entropy of a massive scalar field in the background of the Einstein universe. We determine numerically the structure of the UV-divergent terms. We study analytically the IR term that originates in the long-range correlations arising from the field zero mode on the sphere. We compare with previous results for a scalar field in a dS or AdS background.
}

\begin{document} 
\maketitle
\flushbottom

\section{Introduction} \label{sec:introduction}

The calculation of entanglement entropy from first principles is a challenging task in $3+1$ dimensions, where techniques such as the replica trick cannot be implemented straightforwardly. There are also conceptual problems associated with the ubiquitous ultraviolet (UV) divergences. Infrared (IR) divergences may appear as well, often linked to the existence of normalizable zero modes of the relevant Hamiltonian. On the other hand, the entanglement entropy displays features of profound significance, such as the scaling with the area of the entangling surface for a system in its ground state \cite{Sorkin:1984kjy,Bombelli:1986rw,Srednicki:1993im}, or the link with the $a$-theorem \cite{Casini:2017vbe}, which extends a similar link with the $c$- and $F$-theorems in lower dimensions \cite{Casini:2012ei}. 

Most of the studies of entanglement entropy have been carried out in flat space. In recent years, however, the interest has shifted towards curved backgrounds.\footnote{See, for example, \cite{Shekar:2024edg} for an approach based on the replica trick.} A case of interest is that of de Sitter (dS) space, which is relevant for the inflationary scenario and includes a horizon \cite{Maldacena:2012xp,Boutivas:2024sat,Boutivas:2024lts}. Remarkably, the link with the $a$-theorem can be extended to this background \cite{Abate:2024nyh}. Another case of high interest is that of anti-de Sitter (AdS) space, for which a calculation from first principles has been performed recently by some of the current authors \cite{Boutivas:2025ksp}. That analysis considers spherical regions around the origin of AdS. Therefore, it is not related to the Ryu-Takayanagi conjecture \cite{Ryu:2006bv,Ryu:2006ef} for the holographic entanglement entropy in the context of the AdS/CFT correspondence, for which the entangling surface is located on the AdS boundary. 

For the maximally symmetric spaces in $3+1$ dimensions, the general structure of the entanglement entropy of a free, massive, scalar field in its vacuum state, with a spherical entangling surface of proper area $A$, has the form
\begin{equation} \label{eq:genentropy}
	S_{\textrm{EE}}=\frac{c_1}{4\pi}\frac{A}{\epsilon^2}+
	\left(c_2 + c_3 \frac{A}{4\pi} \mu^2 + c_4 \frac{A}{4\pi a^2} \right)\ln\frac{a}{\epsilon} + 
	c_{\rm IR} \frac{A}{4\pi a^2} \ln\frac{L}{a} +
	\textrm{finite},
\end{equation}
where $a$ is a length scale linked to the curvature of the background, $\epsilon$ is the UV cutoff, $L$ the size of the overall system, and $\mu$ the mass of the field. For dS space $a=1/H$, with $H$ the Hubble constant, while for AdS space $a$ is the AdS length $a_{\rm AdS}$. The first term quantifies the leading UV divergence, which has the same form in all backgrounds. The exact value of $c_1$ depends on the choice of the UV regulator $\epsilon$. This term displays the typical dependence of the entropy on the area of the entangling surface for a theory in its ground state. The coefficients $c_2$, $c_3$, and $c_4$ in the subleading UV-divergent term are independent of the choice of regulator. The value of $c_2$ is linked to the conformal anomaly and is known to be equal to $-1/90$ \cite{Solodukhin:2008dh,Casini:2010kt,Casini:2009sr,Lohmayer:2009sq}. The coefficient $c_3$ was computed in  \cite{Boutivas:2025ksp} for AdS space, where it was found to take the value $-1/6$, in agreement with the expectation from \cite{Hertzberg:2010uv}. The coefficient $c_4$ was found to take the value $1/3$ in dS space \cite{Boutivas:2024lts} and $-1/3$ in AdS space \cite{Boutivas:2025ksp}. These two values are related through the analytic continuation $H^2\to -1/a^2_{\rm AdS}$. It is also noteworthy that the values of $c_3$ and $c_4$ are such that the corresponding contributions to the logarithmically divergent term cancel for a conformally coupled scalar. This is in agreement with the result of \cite{Solodukhin:2008dh} for the entanglement entropy of a conformal theory in a general gravitational background.

The UV-finite terms are more difficult to compute with the method we employ, which is based on the original approach of Srednicki \cite{Srednicki:1993im}, because of the dominance of the divergent terms. Deducing the general structure of the finite part requires precision in the numerical part of the calculation beyond our current capability. It has been possible, however, to identify the presence of an interesting IR term with an unexpected structure, displayed in \eqref{eq:genentropy}. This term appears only for a massless field in dS space in the Bunch-Davies vacuum. It exhibits a dependence on the size $L$ of the overall system and not on the subsystem under consideration. It was shown, both numerically \cite{Boutivas:2024lts} and analytically \cite{Boutivas:2024sat}, that $c_{\rm IR}=1/3$. In that calculation, the extent of the overall system was limited by imposing Dirichlet boundary conditions with vanishing field at a radius $L$. It must be noted that this term is always subleading to the dominant UV-divergent term within the range of validity of the analysis of \cite{Boutivas:2024sat,Boutivas:2024lts}. As a result, it does not affect the scaling properties of the entanglement entropy, which is bounded by the number of degrees of freedom of the smaller subsystem \cite{Page:1993df}. However, it indicates a dependence on the overall system size, even when $L$ exceeds the Hubble radius $1/H$. This behavior is linked to the appearance of an IR divergence in the massless scalar theory in the Bunch-Davies state \cite{Allen:1987tz,Allen:1985ux}, in which the theory was assumed to be lying \cite{Boutivas:2024sat,Boutivas:2024lts}. The IR term in the entanglement entropy results from the extreme squeezing of the wave functions of the IR modes of the theory. Such a term is absent in flat or AdS spaces.

In the current work, we apply our method to a simpler case, i.e., the Einstein universe. We consider a free, massive, scalar scalar field in the $\mathbb{R}\times$S$^d$ background
\begin{equation}\label{eq:metricsphere}
	ds^2=-dt^2+dw^2+a^2\sin^2\frac{w}{a} d\Omega_{d-1},
\end{equation}
where $0\leq w \leq a\pi$. Apart from testing the validity of our approach in a different background, we have two specific points in mind:
\begin{itemize}
\item In the limit of vanishing field mass, the spectrum includes a normalizable zero mode. We expect a term analogous to the one proportional to $c_{\rm IR}$ in \eqref{eq:genentropy} to appear in the entropy, similar to the one emerging in dS case in planar coordinates studied in \cite{Boutivas:2024sat,Boutivas:2024lts}. We want to apply our regularization procedure to this new background and compare the results with equation \eqref{eq:genentropy}. This will confirm the connection of the IR term in the entropy with the presence of long-range correlations resulting from the IR sector of the theory.
\item The metric of AdS space in global coordinates differs from \eqref{eq:metricsphere} by a conformal factor $1/\cos^{2}\frac{w}{a}$. The metric can also be expressed in terms of a radial coordinate $r$ related to $w$ through $r=a\tan\frac{w}{a}$. The UV-regularization can be implemented through the discretization of either coordinate. However, the choice drastically affects the leading divergence, which is regulator-dependent. The numerical analysis through the discretization of $w$ shows that the dominant divergent term is of the form \cite{Boutivas:2025ksp}
\begin{equation}
	S^{(2)}= c_1\sin^2\frac{w_R}{a}\,\frac{a^2}{\epsilon^2},
\end{equation}
with $w_R$ corresponding to the radius of the entangling surface. At first sight, there seems to be a disagreement with \eqref{eq:genentropy}, as the proper area of the entangling surface is $A=4\pi a^2\tan^2{\frac{w_R}{a}}$ for AdS space. This point was discussed in detail in \cite{Boutivas:2025ksp}, where it was argued that there is a relative factor of $1/\cos^{2}{\frac{w}{a}}$ for the number of degrees of freedom in the discretization of $r$ relative to the discretization of $w$. When this is taken into account, agreement with \eqref{eq:genentropy} is achieved. Moreover, the correct dependence on the proper area is reproduced automatically by the analysis of the coefficient of the term proportional to $ \ln \frac{a}{\epsilon}$, which is independent of the choice of the regulator. We want to revisit this point in the context of the background \eqref{eq:metricsphere} to confirm the above conclusions through the computation of the terms proportional to $c_1$ and $c_2$.
\end{itemize}
 
The structure of the paper is as follows: In section \ref{sec:Field_in_S_d}, we formulate the theory of a free scalar on the $\mathbb{R}\times$S$^d$ background and discuss the mode functions. In section \ref{sec:EntEnt} we introduce the discretized version of the theory and summarize the procedure for computing the entanglement entropy in the ground state. In section \ref{sec:numresults}, we describe the numerical method and present the result for the UV-divergent terms in the entropy. In section \ref{sec:analytic}, we present a detailed analytical calculation of the infrared term in the entropy. In section \ref{sec:discussion}, we summarize our findings and compare them with previous results for a scalar field in a dS or AdS background. In the two appendices, we present some expressions that are relevant for the analytical calculation. 

\section{Free Scalar Field in $\mathbb{R}\times$S$^d$ Geometry}
\label{sec:Field_in_S_d}
In this section, we summarize the basic features of the theory of a free scalar field in the $\mathbb{R}\times$S$^d$ background \eqref{eq:metricsphere}. We discuss the action, the admissible boundary conditions, and the eigenfunctions of the corresponding Laplace operator. The parameter $a$ is the radius of the $d$-dimensional sphere. In the decompactification limit $a\rightarrow\infty$ with every other scale kept fixed, $w$ becomes the radial coordinate in the spherical coordinate system that foliates the resulting flat space. The action of a scalar field in this background reads
\begin{equation}
	S=\frac{1}{2}\int dt \int_0^{a\frac{\pi}{2}} dw\int_{\textrm{S}^{d-1}} d\Omega_{d-1}\left(a\sin\frac{w}{a}\right)^{d-1}\left[\dot{\phi}^2-\left(\partial_w\phi\right)^2+\frac{\phi\Delta_{d-1}\phi}{a^2\sin^2\frac{w}{a}}-\mu^2\phi^2\right],
\end{equation}
where $\Delta_{d-1}$ is the Laplacian on the $(d-1)$-dimensional sphere of unit radius. The dot denotes the derivative with respect to time, and $\mu$ is the mass of the field. 

We are interested in calculating the entanglement entropy when the entangling surface lies at a fixed value of $w$. Therefore, we exploit the residual spherical symmetry to expand the field into modes as
\begin{equation}\label{eq:mode_expansion_sphere}
	\phi\left(t,w,\hat{r}\right)=\frac{1}{a^{\frac{d-1}{2}}\sin^{\frac{d-1}{2}}\frac{w}{a}}\sum_{\ell,\vec{m}}\phi_{\ell\vec{m}}\left(t,w\right)Y_{\ell\vec{m}}\left(\hat{r}\right),
\end{equation}
where $\vec{m}=(m_1,\dots,m_{d-2})$ and $\hat{r}$ is a unit vector. The real hyper-spherical harmonics $Y_{\ell\vec{m}}\left(\hat{r}\right)$ obey
\begin{equation}
	\Delta_{d-1}Y_{\ell\vec{m}}\left(\hat{r}\right)=-\ell\left(\ell+d-2\right)Y_{\ell\vec{m}}\left(\hat{r}\right)
\end{equation}
and the orthogonality condition
\begin{equation}
	\int_{\textrm{S}^{d-1}}d\Omega_{d-1}\,Y_{\ell\vec{m}}\left(\hat{r}\right)Y_{\ell\vec{m}^\prime}\left(\hat{r}\right)=\delta_{\ell\ell^\prime}\delta_{\vec{m}\vec{m}^\prime}.
\end{equation}
The action assumes the form 
\begin{equation}
	S=\sum_{\ell,\vec{m}}S_{\ell\vec{m}},
\end{equation}
where the action describing the degrees of freedom of each angular-momentum sector reads
\begin{multline}\label{eq:action_ell_m}
	S_{\ell\vec{m}}=\frac{1}{2}\int dt \int_0^{a\pi} dw \left[\dot{\phi}_{\ell\vec{m}}^2-\left(\partial_w\phi_{\ell\vec{m}}\right)^2-\frac{\nu^2-\frac{1}{4}}{a^2\sin^2\frac{w}{a}}\phi_{\ell\vec{m}}^2-\tilde{\mu}^2\phi^2_{\ell\vec{m}}\right]\\
	+\frac{d-1}{4a}\int dt \int_0^{a\pi} dw \partial_w\left[\cot\frac{w}{a}\phi_{\ell\vec{m}}^2\right],
\end{multline}
with the parameters $\tilde{\mu}^2$ and $\nu$ given by
\begin{equation}\label{eq:mu_tilde_and_nu}
	\tilde{\mu}^2=\mu^2-\frac{\left(d-1\right)^2}{4a^2},\qquad \nu=\ell+\frac{d}{2}-1.
\end{equation}
A field that is coupled non-minimally with the background via a term $-\frac{1}{2}\xi {\cal R} \phi^2$ develops an effective mass term
\begin{equation}
	\mu_{\textrm{eff}}^2=\tilde{\mu}^2+\xi\frac{d(d-1)}{a^2}.
\end{equation}
Thus, for $\xi=\frac{d-1}{4d}$ we have $\mu^2_{\textrm{eff}}=\mu^2$.  For $\mu^2=0$, the theory is Weyl invariant.

In order to specify the admissible boundary conditions, we vary the action with respect to $\phi_{\ell\vec{m}}$ obtaining
\begin{multline}
	\delta S_{\ell\vec{m}}=\int dt \int_0^{a\pi} dw \left[-\ddot{\phi}_{\ell\vec{m}}+\partial^2_w\phi_{\ell\vec{m}}-\frac{\nu^2-\frac{1}{4}}{a^2\sin^2\frac{w}{a}}\phi_{\ell\vec{m}}-\tilde{\mu}^2\phi_{\ell\vec{m}}\right]\delta\phi_{\ell\vec{m}}\\
	-\int dt \int_0^{a\pi} dw \partial_w\left[\delta\phi_{\ell\vec{m}}\sin^{\frac{d-1}{2}}\frac{w}{a}\,\partial_w\frac{\phi_{\ell\vec{m}}}{\sin^{\frac{d-1}{2}}\frac{w}{a}}\right].
\end{multline}
The mode functions are calculated using the solutions of the eigensystem
\begin{equation}\label{eq:eigenvalue_prob_cont}
	-\partial^2_w\Phi+\frac{\nu^2-\frac{1}{4}}{a^2\sin^2\frac{w}{a}}\Phi=\left(\omega^2-\tilde{\mu}^2\right)\Phi,
\end{equation}
which can also be written in the form
\begin{equation}\label{eq:eigenvalue_prob_cont_theta}
	-\partial^2_\theta\Phi+\frac{\nu^2-\frac{1}{4}}{\sin^2\theta}\Phi=E\Phi,
\end{equation}
where $\theta=\frac{w}{a}$ and $E=a^2\left(\omega^2-\tilde{\mu}^2\right)$. For Dirichlet boundary conditions 
the solution reads 
\begin{equation}\label{eq:eigenfunctions}
	\Phi_{n\ell}(\theta)=c_n\sqrt{\sin\theta} P_{n+\nu}^{-\nu}(\cos\theta),\qquad E_n=\left(n+\nu+\frac{1}{2}\right)^2,
\end{equation}
where $P_{n+\nu}^{-\nu}$ are associated Legendre polynomials with $n\in\mathbb{N}$. The appropriate normalization factor equals
\begin{equation}
	c_n=\sqrt{n+\nu+\frac{1}{2}}\sqrt{\frac{(n+2\nu)!}{n!}}.
\end{equation}
The eigenfrequencies $\omega_n$ read
\begin{equation}
	\omega_n=\frac{1}{a}\left[\tilde{\mu}^2a^2+\left(n+\nu+\frac{1}{2}\right)^2\right]^{1/2}.
\end{equation}
The eigenfunctions are orthonormal
\begin{equation}
	\int_0^\pi d\theta\, \Phi_{n\ell}(\theta)\Phi_{m\ell}(\theta)=\delta_{n,m},
\end{equation} 
and satisfy the completeness relation\footnote{The eigenfunctions can also be written in terms of the Gegenbauer polynomials $C_n^{\nu}(z)$ as
\begin{equation}
	\Phi_{n\ell}(\theta)=c^\prime_n \sin^{\nu+\frac{1}{2}}\theta\, C_n^{\nu+\frac{1}{2}}(\cos\theta),\qquad c^\prime_n=\frac{\sqrt{n+\nu+\frac{1}{2}}}{\sqrt{\pi}}\sqrt{\frac{n!}{(n+2\nu)!}}2^\nu \Gamma\left(\nu+\frac{1}{2}\right).
\end{equation}
The completeness relation follows from the summation formula
\begin{equation}
	\frac{2^{2 \lambda -1} \Gamma (\lambda )^2}{\left(1-x^2\right)^{\frac{1}{4} (1-2 \lambda )}\left(1-y^2\right)^{\frac{1}{4} (1-2 \lambda )}}\sum _{n=0}^{\infty } \frac{n! (\lambda +n) C_n^{(\lambda )}(x) C_n^{(\lambda )}(y)}{\Gamma (n+2 \lambda)}=\pi \delta (x-y),
\end{equation}
which holds for $\Re(\lambda )>-\frac{1}{2}$, $\lambda \neq 0$, $ -1<x<1$ and $-1<y<1$.}
\begin{equation}
	\sum_{n=0}^{\infty}\Phi_{n\ell}(\theta)\Phi_{n\ell}(\theta^\prime)=\delta(\theta-\theta^\prime).
\end{equation}

Strictly speaking, since the potential of the effective Schr\"odinger equation \eqref{eq:eigenvalue_prob_cont_theta} is divergent at $\theta=0$ and $\theta=\pi$, we have imposed regularity conditions at the endpoints, rather than boundary ones. Moreover, the conditions imposed on each $(1+1)$-dimensional problem with integer $\ell$ should enforce the consistency of the original $(d+1)$-dimensional theory with $d\geq 2$. The expansion \eqref{eq:mode_expansion_sphere} of the original field includes a factor $\sin^{\frac{1-d}{2}}\theta$ that diverges at $\theta=0$ and $\pi$. Even though a general solution of equation \eqref{eq:eigenvalue_prob_cont_theta} can scale as $\theta^{\frac{1}{2}\pm \nu}$ and $\left(\pi-\theta\right)^{\frac{1}{2}\pm \nu}$ near the endpoints, only the exponent $\frac{1}{2} + \nu$ corresponds to regular, normalizable modes. These are given by \eqref{eq:eigenfunctions} with $n\in\mathbb{N}$, for which
\begin{equation}
	\frac{\Phi_{n\ell}(\theta)}{\sin^{\frac{d-1}{2}}\theta}\propto \theta^{\ell}
\end{equation}
for $\theta\rightarrow0$, and correspondingly for $\theta\rightarrow\pi$. As a result, the eigenfunctions \eqref{eq:eigenfunctions} are the only ones that lead to field configurations that do not diverge. All contributions to the original field  $\phi\left(t,w,\hat{r}\right)$ from angular-momentum sectors with $\ell\geq 1$ obey Dirichlet boundary conditions, whereas the $\ell=0$ sector leads to Neumann conditions. These regularity conditions guarantee that the original field is smooth all over the sphere S$^d$ and no conical singularity is introduced.

Before proceeding to the main part of the calculation, let us examine the lowest mode of the spectrum of each angular-momentum sector. For $n=0$, using the specific form of the Legendre function, namely $P_{\nu }^{-\nu }(z)=2^{-\nu }\left(1-z^2\right)^{\nu /2}/ \Gamma (\nu +1)$, we obtain
\begin{equation}\label{eq:zero_mode}
	\Phi_{0\ell}(\theta)=\sqrt{\frac{\Gamma\left(\nu+\frac{3}{2}\right)}{\sqrt{\pi}\Gamma\left(\nu+1\right)}}\sin^{\nu+\frac{1}{2}}\theta,\qquad \omega_0=\sqrt{\mu^2+\frac{\ell\left(\ell+d-1\right)}{a^2}}.
\end{equation}
For $\ell=0$ and $\mu=0$, the spectrum contains a zero mode in any number of dimensions. This mode corresponds to a
configuration of the original field that is constant and obeys Neumann conditions at both endpoints $\theta=0$ and $\theta=\pi$.

\section{Ground State Entanglement Entropy}
\label{sec:EntEnt}
\subsection{Discretization}
The calculation of entanglement entropy is performed through the discretization of the system \cite{Sorkin:1984kjy,Bombelli:1986rw,Srednicki:1993im}. For this purpose, we utilize the Hamiltonian that corresponds to 
the action \eqref{eq:action_ell_m}, which reads
\begin{equation}\label{eq:Hamiltonian_ell_m}
	H_{\ell\vec{m}}=\frac{1}{2}\int_0^{a\pi} dw \left[\dot{\phi}_{\ell\vec{m}}^2+\left(\partial_w\phi_{\ell\vec{m}}\right)^2+\frac{\nu^2-\frac{1}{4}}{a^2\sin^2\frac{w}{a}}\phi_{\ell\vec{m}}^2+\tilde{\mu}^2\phi^2_{\ell\vec{m}}\right].
\end{equation}
For $a\gg w$, the Hamiltonian reduces to that of a massive field in flat space. 

We employ the discretization scheme introduced in \cite{Boutivas:2025ksp}, namely
\begin{equation}\label{eq:discretization_1}
	w=\epsilon\,i,\qquad a=\frac{1}{\pi}\epsilon\left(N+1\right),\qquad \int_{0}^{a\pi}dw\rightarrow\epsilon\sum_{i=0}^{N+1},
\end{equation}
with the discrete modes given by
\begin{equation}\label{eq:discretization_2}
	\phi_{\ell\vec{m}}\left(t,w\right)\rightarrow\frac{1}{\sqrt{\epsilon}}\phi_{\ell\vec{m},i}\left(t\right),\qquad \pi_{\ell\vec{m}}\left(t,w\right)\rightarrow\frac{1}{\sqrt{\epsilon}}\pi_{\ell\vec{m},i}\left(t\right).
\end{equation}
The Hamiltonian of the discrete system reads
\begin{equation}
	H_{\ell\vec{m}}=\frac{1}{2}\sum_{i}\left[\pi^2_{\ell\vec{m},i} +\frac{\left(\phi_{\ell\vec{m},i+1}-\phi_{\ell\vec{m},i}\right)^2}{\epsilon^2}+\left(\frac{\nu^2-\frac{1}{4}}{a^2\sin^2\frac{\pi i}{N+1}}+\tilde{\mu}^2\right)\phi^2_{\ell\vec{m},i}\right],
\end{equation}
where $a$ is given by \eqref{eq:discretization_1}. We implement Dirichlet boundary conditions by imposing $\phi_{\ell\vec{m},0}(t)=\phi_{\ell\vec{m},N+1}(t)=0$ and $\pi_{\ell\vec{m},0}(t)=\pi_{\ell\vec{m},N+1}(t)=0$. Thus, we obtain a system of $N$ coupled harmonic oscillators, whose dynamics is governed by the Hamiltonian
\begin{equation}\label{eq:H_ell_m}
	H_{\ell\vec{m}}=\frac{1}{2}\sum_{i=1}^{N}\pi^2_{\ell\vec{m},i}+\frac{1}{2}\sum_{i,j=1}^{N}\phi_{\ell\vec{m},i}K_{ij}\phi_{\ell\vec{m},j},
\end{equation}
with elements of the coupling matrix $K_{ij}$ given by
\begin{equation}\label{eq:coupling_matrix_1}
	K_{ij}=\frac{1}{\epsilon^2}\left[\left(2+\frac{\pi^2}{(N+1)^2}\frac{\nu^2-\frac{1}{4}}{\sin^2\frac{\pi i}{N+1}}+\tilde{\mu}^2\epsilon^2\right)\delta_{i,j}-\delta_{i+1,j}-\delta_{i,j+1}\right].
\end{equation}
We note that $\tilde{\mu}$ depends on $a$, see equation \eqref{eq:mu_tilde_and_nu}. 

The above form of $K$ is suitable for calculations in which dimensionful quantities are measured in units of $\epsilon$,
which is equivalent to setting $\epsilon=1$. For our calculation it is more convenient to express dimensionful quantities in terms 
of the curvature scale $a$. 
Thus, we write the coupling matrix in the form
\begin{equation}\label{eq:coupling_matrix_2}
	K_{ij}=\frac{1}{a^2}\left[\frac{(N+1)^2}{\pi^2}\left(2\delta_{i,j}-\delta_{i+1,j}-\delta_{i,j+1}\right)+\left(\frac{\nu^2-\frac{1}{4}}{\sin^2\frac{\pi i}{N+1}}+\tilde{\mu}^2a^2\right)\delta_{i,j}\right]
\end{equation}
and set $a=1$. For growing $N$, with $x=i/(N+1)$ and $y=j/(N+1)$ kept fixed, we approach the continuous theory. 
We have confirmed that the eigenvalues and eigenfunctions of the above matrix for large $N$ agree with the 
solutions (\ref{eq:eigenfunctions}) of the continuous theory. 

\subsection{The Entropy}
\label{subsec:Calculation_Entropy}
The Hamiltonian \eqref{eq:H_ell_m} describes a system of $N$ coupled harmonic oscillators. Hence, one can implement methods of quantum mechanics to calculate the entanglement entropy \cite{Sorkin:1984kjy,Bombelli:1986rw,Srednicki:1993im}. The calculation is practically identical to the pioneering calculation of \cite{Srednicki:1993im}. The only difference is that the coupling matrix corresponds to a field in $\mathbb{R}\times$S$^d$ and not in flat space.

The wave function that describes the modes of each $(\ell,\vec{m})$-sector is given by
\begin{equation}
	\Psi({{\bm{\phi}}_{\ell\vec{m}}})=\left(\det\frac{\Omega}{\pi}\right)^{1/4} e^{-\frac{1}{2} {\bm{\phi}}_{\ell\vec{m}}^T \Omega\, {\bm{\phi}}_{\ell\vec{m}}},
\end{equation}
where $\Omega$ stands for the positive square root of the matrix $K$, and ${\bm{\phi}}_{\ell\vec{m}}$ is the column vector that contains $\phi_{\ell\vec{m},j}$, the values of the field at the various lattice points. Oscillators $1$ to $n$ comprise the subsystem $A$, while oscillators $n+1$ to $N$ comprise the complementary subsystem $C$. By tracing out the subsystem $C$, we can calculate the entanglement entropy of the subsystem $A$, which reads
\begin{equation}\label{eq:SEE_of_M}
	S_{\ell\vec{m}}^{\textrm{EE}}=\sum_{i=1}^{n}\left(\frac{\sqrt{\lambda_i}+1}{2}\ln\frac{\sqrt{\lambda_i}+1}{2}-\frac{\sqrt{\lambda_i}-1}{2}\ln\frac{\sqrt{\lambda_i}-1}{2}\right),
\end{equation}
where $\lambda_i$ are the eigenvalues of the matrix
\begin{equation}\label{eq:Matrix_M}
	\mathcal{M}=\left(\Omega^{-1}\right)_A\left(\Omega\right)_A.
\end{equation}
Here the matrix $\left(\Omega\right)_A$ is the $n\times n$ top-left block of $\Omega$ and similarly $\left(\Omega^{-1}\right)_A$ is the $n\times n$ top-left block of $\Omega^{-1}$. This formula relates the entanglement entropy of a subsystem to the two-point correlation functions of the field and the conjugate momentum restricted to the particular subsystem. This approach was introduced in \cite{Peschel:2002yqj} and is equivalent to that of \cite{Srednicki:1993im}. A review of the method is provided in \cite{Casini:2009sr}; see also \cite{Katsinis:2024gef,Katsinis:2023hqn} for more details about the equivalence mentioned above. We obtain the total entanglement entropy by summing the contributions of all angular-momentum sectors. More specifically, for $d=3$, i.e., for $\mathbb{R}\times$S$^3$, we have
\begin{equation}\label{eq:SEE_total}
	S_{\textrm{EE}}=\sum_{\ell=0}^\infty(2\ell+1)S_{\ell}^{\textrm{EE}}.
\end{equation}

In analogy with \cite{Boutivas:2025ksp}, we must address a technical issue. The indices $0,1,\dots, N,N+1$ label the locations of the degrees of freedom of the discretized system. As a result, the last degree of freedom of the subsystem $A$ is associated with the index $n$. Similarly, the first degree of freedom of the subsystem $C$ is associated with the index $n+1$. The entanglement entropy is located between these nodes. Therefore, we can take
\begin{equation}
	\frac{w_R}{a}=\pi\frac{n+\frac{1}{2}}{N+1},\label{eq:wR}
\end{equation}
where $w_R$ is the location of the entangling surface in the continuous theory. As we work with a fixed sphere radius $a$, incorporating this shift at the level of the coupling matrix improves the convergence to the continuous result.
Thus, instead of using the coupling matrix \eqref{eq:coupling_matrix_1}, we use the coupling matrix resulting from the shift $i\to i+\frac{1}{2}$ in the argument of the sine. Without this trick, a series of spurious terms would have emerged, which would have to be recombined to express the final result as a function of $n+\frac{1}{2}$. The results of the numerical calculation verify that this particular coupling matrix indeed facilitates the convergence to the continuum limit.

\section{Numerical Analysis and Results}
\label{sec:numresults}

\subsection{Methodology}
We perform the numerical computation for a field defined on the background $\mathbb{R} \times \text{S}^3$, corresponding to $d = 3$. Following the approach of \cite{Boutivas:2025ksp}, we aim to investigate how the entanglement entropy depends not only on the radius of the entangling surface and the curvature scale of the background but also on the field's mass $\mu$.

The foundational study \cite{Srednicki:1993im}, which analyzes a massless scalar field in flat space, focuses on the continuum theory in the infinite-size limit. In that framework, fixing the lattice spacing for numerical computations is natural. Formally, the continuum limit is approached by taking $\epsilon \to 0$, $n \to \infty$, and $N \to \infty$, while keeping $R = n\epsilon$ finite and fixed, and allowing $L = N\epsilon \to \infty$. To numerically access this regime, one typically fixes the value of $n$ and varies the total number of degrees of freedom $N$. By extrapolating the results to $N \to \infty$, one isolates the entanglement entropy in the infinite-size limit, where the relevant dimensionless ratio is $N \simeq L/\epsilon$. In this limit, the entanglement entropy becomes a function solely of $n$, the only remaining dimensionless parameter. Since $n \simeq R/\epsilon$, the continuum result for the entanglement entropy corresponds to the same functional dependence, with $n$ effectively replaced by $R/\epsilon$. As a result, when this function is expanded in powers of $n$, only non-negative powers contribute to the continuum limit. Further details are provided in \cite{Lohmayer:2009sq}.

In the present work, the system has finite size---specifically, the interval $[0, a\pi]$. Consequently, fixing the curvature scale $a$ rather than the lattice spacing is more natural. With this choice, equation \eqref{eq:discretization_1} establishes a relationship between the UV cutoff and the number of lattice sites. The continuum limit is then approached by taking $N \to \infty$.

The numerical computations are carried out using custom C++ code built upon the Eigen library for linear algebra. The implementation supports 128-bit arithmetic, enabling a precision of approximately 33 to 35 significant digits.

We briefly outline the setup of the numerical calculation. The discretization scheme specified in equations \eqref{eq:discretization_1} and \eqref{eq:discretization_2} places the degrees of freedom at radii $w=w_i$, given by
\begin{equation}
	\frac{w_i}{a}=\pi\frac{i}{N+1},\qquad i=1,\dots , N .
\end{equation} 
We consider a series of lattices composed of spherical shells with $N = 49 + 50k$, for $k=0,\ldots,8$. For each value of $N$, the entanglement entropy is computed for entangling surfaces located at radii $w_n$, with $n=(k+1)j$ and $j=1,\ldots , 48$, thereby ensuring consistent physical radii across all lattice configurations. This allows for meaningful comparisons of entanglement entropies at different lattice spacings. The computation is repeated for several values of $\mu^2a^2$, specifically,
\begin{equation}
	\mu^2a^2 \in \left\{0.05, 0.10, 0.15, 0.30, 0.40, 0.50, 1.00, 1.50, 2.00, 5.00\right\}.
\end{equation}
The case $\mu^2a^2=1$  corresponds to the conformally coupled scalar field. 

Recall that the total entanglement entropy is given by the sum in equation \eqref{eq:SEE_total}. Since it is unfeasible to compute contributions from infinitely many values of $\ell$, we restrict our calculation to angular-momentum sectors ranging from $\ell = 0$ to $\ell = 4 \cdot 10^4$ and then extrapolate to estimate the value of the full sum.

To build intuition about the numerical results, we briefly discuss some qualitative expectations. Since we study the ground state, the leading divergence is expected to scale as $1/\epsilon^2$. For fixed $a$ in the discrete theory, this implies a leading term proportional to $(N+1)^2$, according to equation \eqref{eq:discretization_1}. Furthermore, because the entanglement entropy is an increasing function of the area of the entangling surface, it reaches its maximum when $w_R/a = \pi/2$.

Following the methodology of \cite{Boutivas:2024lts}, we estimate the entanglement entropy by summing the contributions from all angular-momentum sectors in equation \eqref{eq:SEE_total}. To do this, we consider the truncated sum
\begin{equation}
	S_{\textrm{EE}}(n,N,\mu^2a^2;\ell_{\textrm{max}})=\sum_{\ell=0}^{\ell_{\textrm{max}}}(2\ell+1)S_{\ell}^{\textrm{EE}}(n,N,\mu^2a^2)
\end{equation}
as a function of $\ell_{\textrm{max}}$. This sum is found to obey the expansion
\begin{multline}
	S_{\textrm{EE}}(n,N,\mu^2a^2;\ell_{\textrm{max}}) = S_{\infty}(n,N,\mu^2a^2)\\
	+ \sum_{i=1}^{i_\textrm{max}}\frac{1}{\ell_\textrm{max}^{2i}}\left[a_i(n,N,\mu^2a^2)+b_i(n,N,\mu^2a^2)\ln \ell_\text{max}\right],
\end{multline}
in agreement with the behavior observed in \cite{Srednicki:1993im}. By including a sufficiently large number of subleading terms, we obtain an accurate estimate of the full entanglement entropy,
\begin{equation}
	S_\infty(n,N,\mu^2a^2)=\lim_{\ell_\textrm{max}\rightarrow\infty}S_{\textrm{EE}}(n,N,\mu^2a^2;\ell_\textrm{max}).
\end{equation}

The analysis proceeds by examining the dependence of $S_{\infty}(n, N, \mu^2 a^2)$ on the ratio $\frac{a}{\epsilon} = \frac{N+1}{\pi}$. We find that $S_{\infty}(n, N, \mu^2 a^2)$ admits the decomposition
\begin{equation}
	S_{\infty}(n,N,\mu^2a^2)= \frac{a^2}{\epsilon^2} S^{(2)}(n,\mu^2a^2)+ S_l^{(0)}(n,\mu^2a^2)\ln\frac{a}{\epsilon}+ S^{(0)}(n,\mu^2a^2)+R(n,N,\mu^2a^2),
\end{equation}
where the remainder term $R(n, N, \mu^2 a^2)$ is given by
\begin{equation}
	R(n,N,\mu^2a^2)=\sum_{i=1}^{i_\textrm{max}} \frac{\epsilon^i}{a^i}S^{(-i)}(n,\mu^2a^2)+\ln\frac{a}{\epsilon}\sum_{j=1}^{j_\textrm{max}} \frac{\epsilon^j}{a^j}S_l^{(-j)}(n,\mu^2a^2).
\end{equation}

In the continuous theory, the remainder $R$ vanishes. However, it plays an important role in the discrete analysis, where isolating finite cutoff effects is essential for accurately determining the coefficients $S^{(2)}(n, \mu^2 a^2)$, $S_l^{(0)}(n, \mu^2 a^2)$, and $S^{(0)}(n, \mu^2 a^2)$. We emphasize that the first two terms, which are associated with the UV-divergent part of the entanglement entropy, are computed with high precision. In contrast, the coefficient $S^{(0)}(n, \mu^2 a^2)$, representing the UV-finite contribution, is more sensitive to numerical error, both from the accumulation of fitting uncertainties and from the limits of numerical precision.

In the next section, we analyze the dependence of the first two terms on the entangling surface radius $w_R$.

\subsection{Results}
\label{subsec:results}
\begin{figure}[t]
	\centering
	\begin{picture}(96.5,60.5)
		\put(2,1){\includegraphics[angle=0,width=0.9\textwidth]{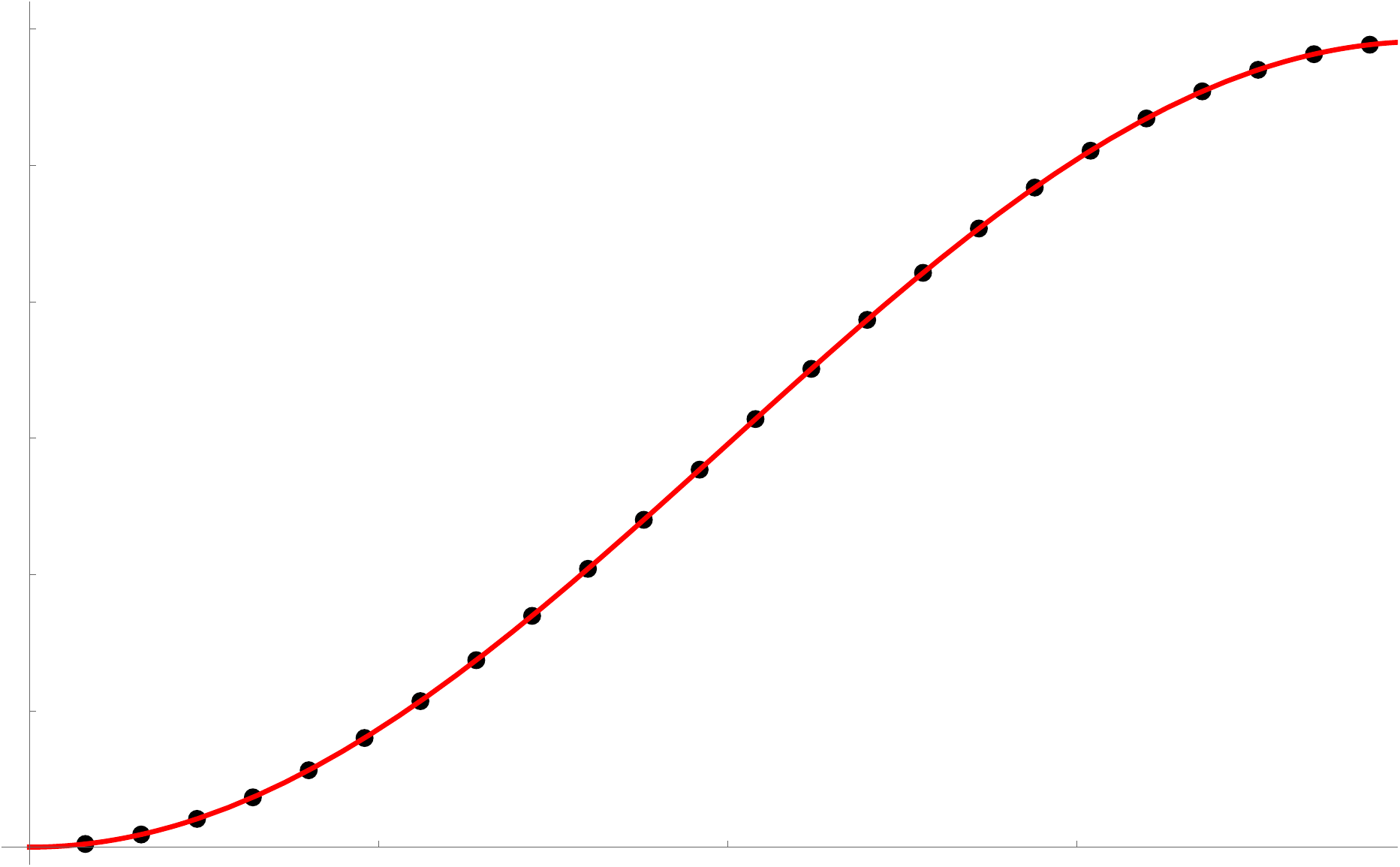}}
		\put(2.4,57){{\Large $S^{(2)}$}}
		\put(92.5,1.4){{\Large $\frac{w_R}{a}$}}
		\put(48,0){{\small $\frac{\pi}{4}$}}
		\put(25.5,0){{\small $\frac{\pi}{8}$}}
		\put(70,0){{\small $\frac{3\pi}{8}$}}
		\put(91,0){{\small $\frac{\pi}{2}$}}
		\put(0,19){{\small $0.1$}}
		\put(0,36.5){{\small $0.2$}}
		\put(0,54.2){{\small $0.3$}}
	\end{picture}
	\caption{Numerical fit for the determination of $S^{(2)}$ as a function of $w_R/a$ for $\mu^2a^2=0$. The data indicate that $S^{(2)}=c_1\sin^2\frac{w_R}{a}$, with $c_1\simeq0.29543145$ for all values of $\mu^2a^2$.}
	\label{fig:area_law}
\end{figure}
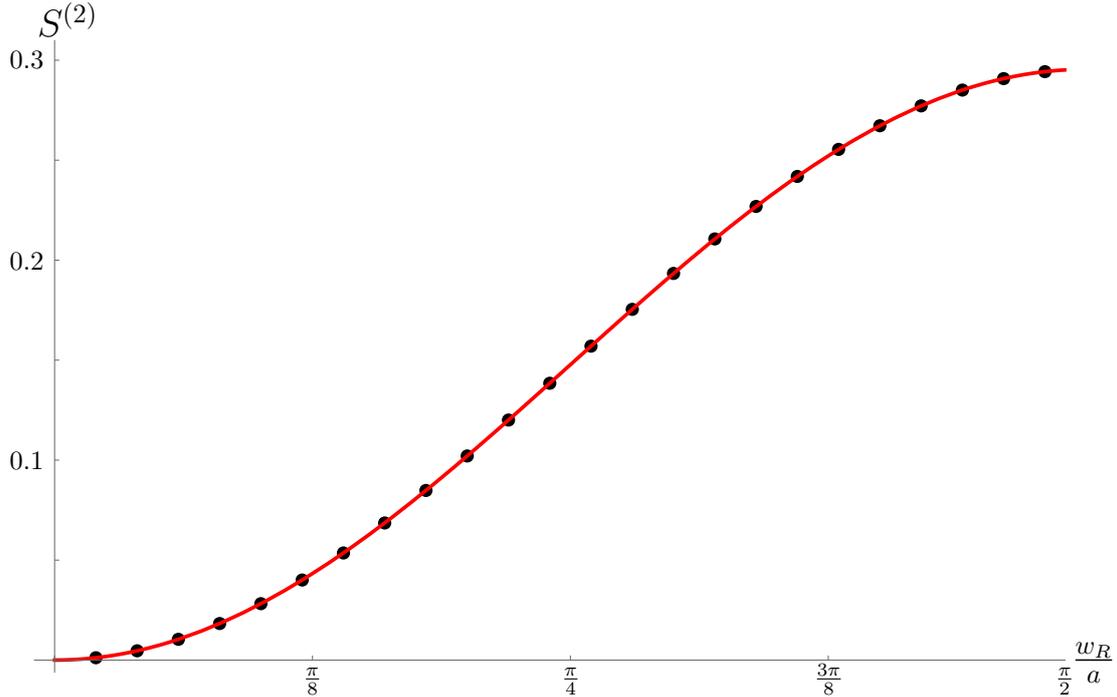

The numerical analysis reveals that $S^{(2)}(n)$, the coefficient of the leading divergence, takes the form
\begin{equation}\label{eq:square_term}
		S^{(2)}(n)=  c_1\sin^2\frac{w_R}{a},
\end{equation}
where the constant $c_1$ is determined through numerical fitting to be
\begin{equation}\label{eq:d1_value}
		c_1\simeq 0.29543145.
\end{equation}
This functional form appears to be a universal feature independent of the mass parameter. Figure~\ref{fig:area_law} illustrates the excellent agreement between the data and the expression in equation~\eqref{eq:square_term} for $\mu^2 a^2 = 0$. Repeating the comparison for other mass values shows no appreciable deviation from this behavior.

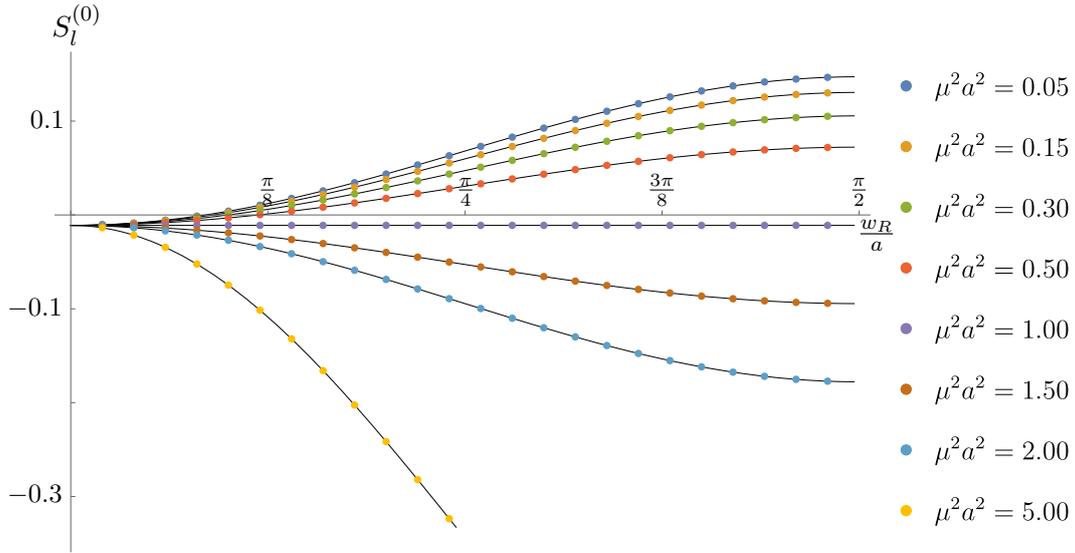
\begin{figure}[t]
	\centering
	\begin{picture}(93.8,49)
		\put(3.8,0){\includegraphics[angle=0,width=0.9\textwidth]{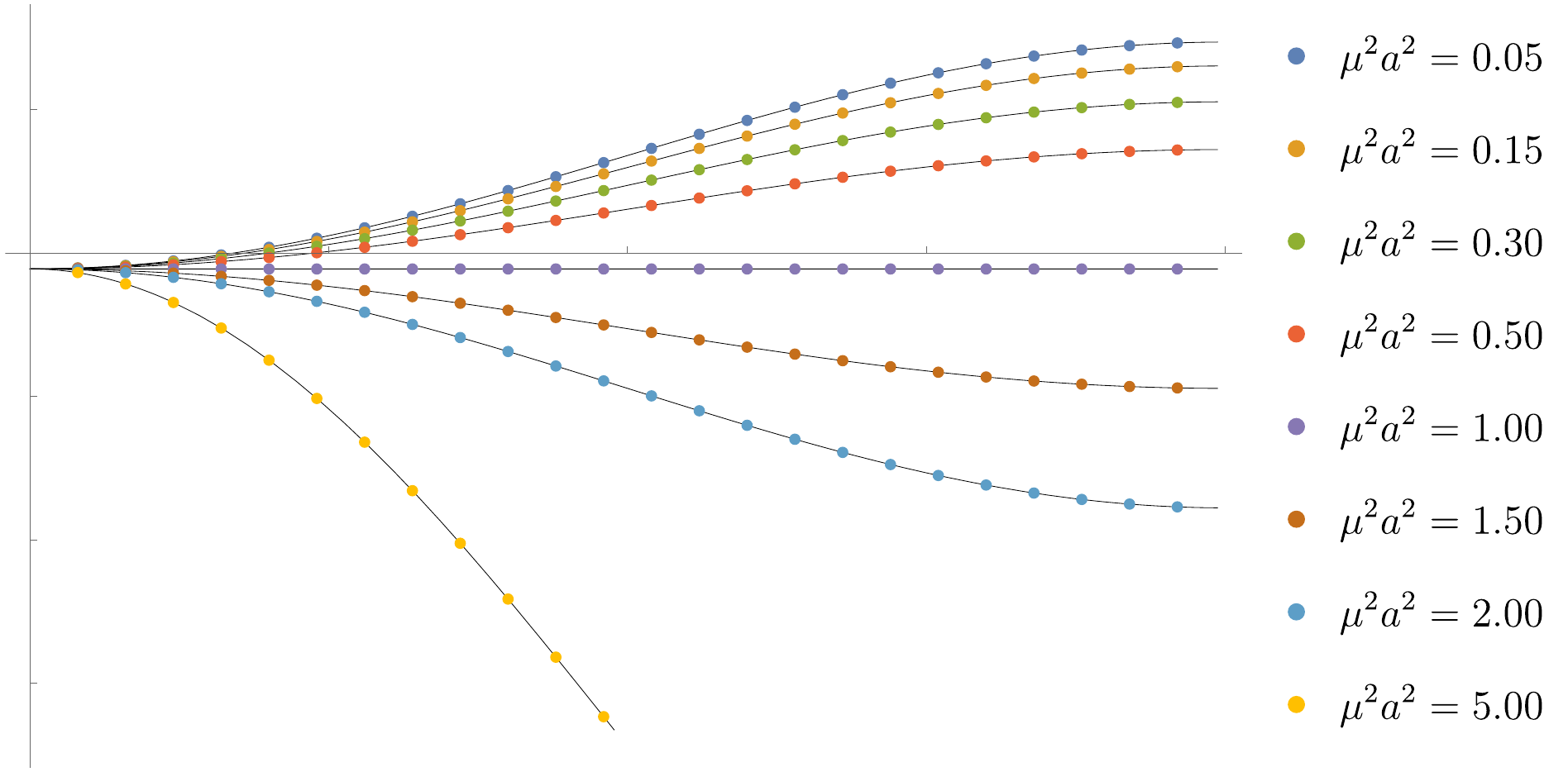}}
		\put(3.9,45.5){$S_l^{(0)}$}
		\put(74,27.5){$\frac{w_R}{a}$}
		\put(21.8,31.4){{\small $\frac{\pi}{8}$}}
		\put(39.05,31.4){{\small $\frac{\pi}{4}$}}
		\put(55.675,31.4){{\small $\frac{3\pi}{8}$}}
		\put(73.3,31.4){{\small $\frac{\pi}{2}$}}
		\put(0,4.6){{\small $-0.3$}}
		\put(0,21.0){{\small $-0.1$}}
		\put(2,37.4){{\small $0.1$}}
	\end{picture}
	\caption{Numerical fits for the determination of $S^{(0)}_l$ as a function of $w_R/a$ for various values of $\mu^2a^2$. The data indicate with high precision that $S^{(0)}_l=a_l(\mu^2a^2)+b_l(\mu^2a^2)\sin^2\frac{w_R}{a}$.}
	\label{fig:log_term}
\end{figure}

It is worth emphasizing that for $w_R \ll a$ equation \eqref{eq:square_term} and the value of $c_1$ correctly reproduce the leading UV divergence in flat space, as obtained in \cite{Srednicki:1993im}. While this term is regulator dependent, the agreement in the value of $c_1$ arises because the leading divergence is a consequence of local effects, and the curved space is locally flat. A similar agreement has also been observed for de Sitter \cite{Boutivas:2024lts} and anti-de Sitter backgrounds \cite{Boutivas:2025ksp}. For finite $w_R / a$, the product $(a^2/\epsilon^2)S^{(2)}$ corresponds to the first term of \eqref{eq:genentropy} with the proper area given by $A=4\pi a^2\sin^2{\frac{w_R}{a}}$.

The result \eqref{eq:square_term} is identical to the one obtained in \cite{Boutivas:2025ksp} for the entanglement entropy in AdS space described in terms of global coordinates, even though the proper area is ${A=4\pi a^2\tan^2{\frac{w_R}{a}}}$ in the AdS case. As we discussed in the introduction, this is a consequence of the strong dependence of this term on the choice of the discretized radial coordinate, which drastically affects the number of degrees of freedom of the system. On the other hand, the regulator-independent term $S^{(0)}_\ell$ displays the correct dependence on the proper area. The form of this term provides a crucial test for the consistency of our approach for the current case, in which the proper area is  $A=4\pi a^2\sin^2{\frac{w_R}{a}}$.

\begin{figure}[t]
	\centering
	\begin{picture}(98.5,29)
		\put(3,1.5){\includegraphics[width=0.4\textwidth]{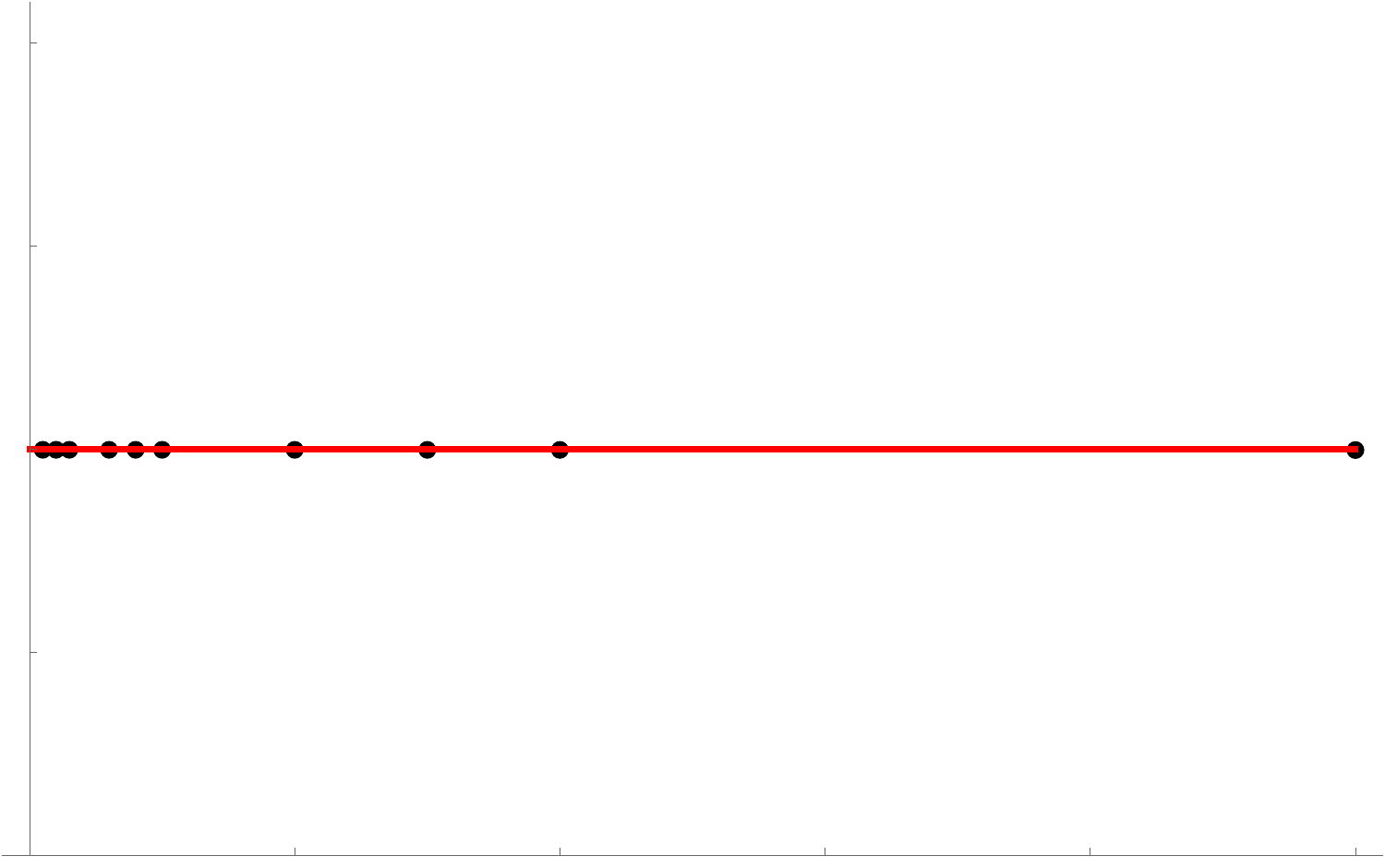}}
		\put(53,0){\includegraphics[width=0.4\textwidth]{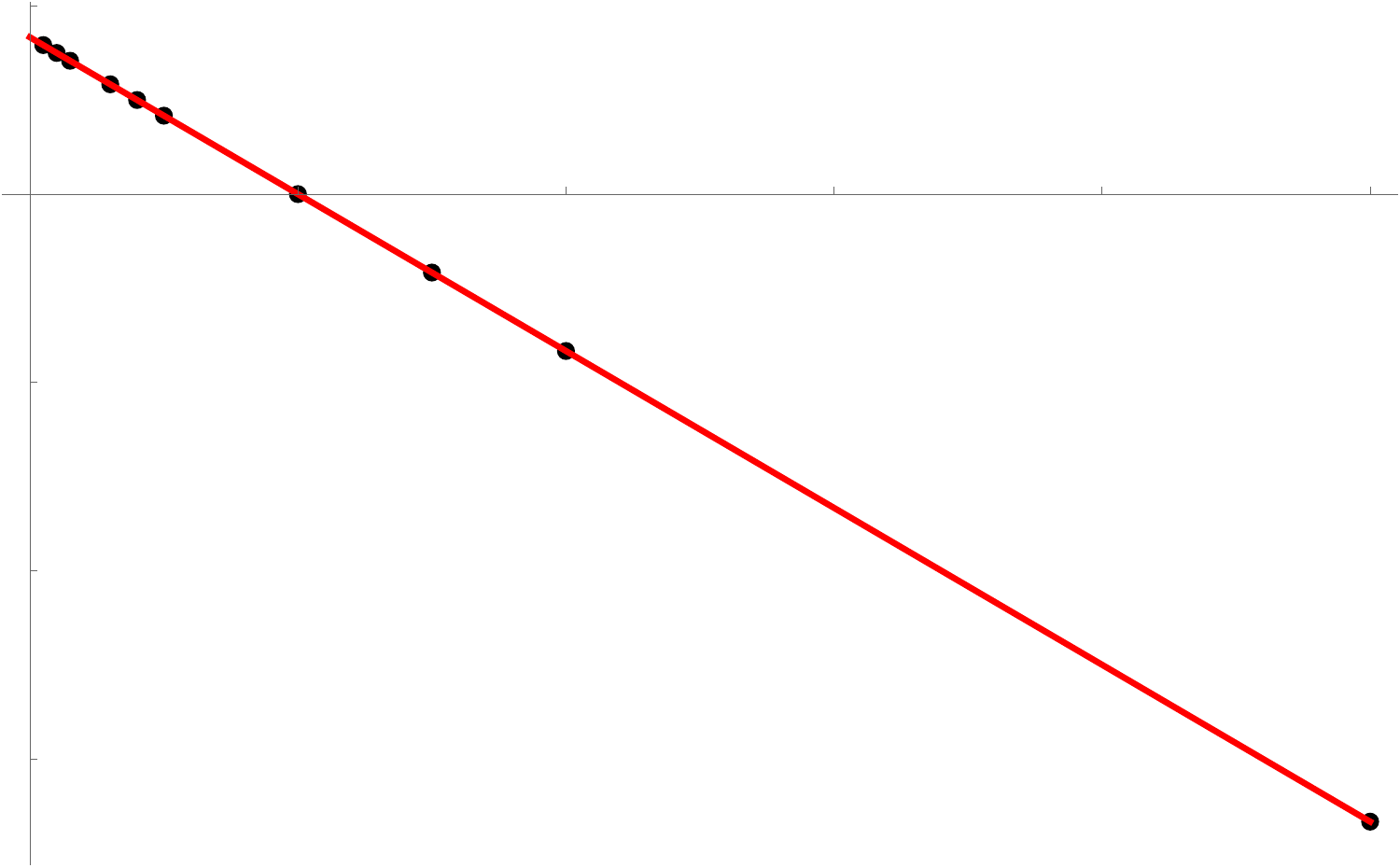}}
		\put(0,27){$-90\,a_l$}
		\put(51.2,26.7){$-b_l$}
		\put(43.5,1.5){$\mu^2a^2$}
		\put(93.5,19){$\mu^2a^2$}
		\put(3.3,0){{\footnotesize $0$}}
		\put(11,0){{\footnotesize $1$}}
		\put(18.7,0){{\footnotesize $2$}}
		\put(26.3,0){{\footnotesize $3$}}
		\put(34.0,0){{\footnotesize $4$}}
		\put(41.6,0){{\footnotesize $5$}}
		\put(61,17.5){{\footnotesize $1$}}
		\put(68.6,17.5){{\footnotesize $2$}}
		\put(76.4,17.5){{\footnotesize $3$}}
		\put(84,17.5){{\footnotesize $4$}}
		\put(91.6,17.5){{\footnotesize $5$}}
		\put(2,13){{\footnotesize $1$}}
		\put(2,24.7){{\footnotesize $2$}}
		\put(49.0,2.5){{\footnotesize $-0.6$}}
		\put(49,7.8){{\footnotesize $-0.4$}}
		\put(49,13.2){{\footnotesize $-0.2$}}
		\put(50.7,24.1){{\footnotesize $0.2$}}
	\end{picture}
	\caption{The coefficients $a_l$ and $b_l$ for various values of $\mu^2a^2$. The left plot demonstrates that $a_l=-1/90$, independently of $\mu^2a^2$, with high precision. The right plot demonstrates that $b_l$ has a linear dependence on $\mu^2a^2$ with a slope equal to $-1/5.9998$ and $y-$intercept equal to $1/6.0009$. We can confidently deduce that $b_l=-\mu^2a^2/6+1/6$.}
	\label{fig:log_term_coef}
\end{figure}
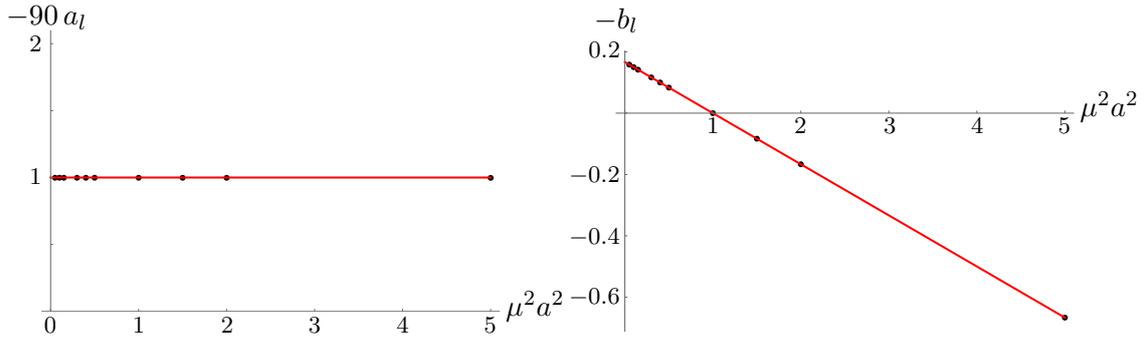
Regarding the coefficient of $\ln\epsilon$, the numerical analysis indicates that it has the form
\begin{equation}
		S_l^{(0)}(n,\mu^2a^2)=a_l(\mu^2a^2) +b_l(\mu^2a^2)\sin^2\frac{w_R}{a}.
\end{equation}
Figure~\ref{fig:log_term} shows excellent agreement between the ansatz and the data across various values of $\mu^2a^2$. Notably, the dependence on $w_R$ appears exclusively though the factor $\sin^2\frac{w_R}{a}$, which is closely related to the proper area of the entangling surface. This confirms that the functional form of the coefficient of the logarithmic term is the expected one in $\mathbb{R}\times$S$^{3}$, similarly to the AdS case \cite{Boutivas:2025ksp}.

The functions $a_l(\mu^2a^2)$ and $b_l(\mu^2a^2)$ can both be fitted using simple expressions. Specifically, we find:
\begin{align}
	a_l=&c_2={\rm const.}, \label{eq:ald2} \\
	b_l(\mu^2a^2)=&c_3\,\mu^2a^2+c_4.	\label{eq:bld4d5}
\end{align}
Figure~\ref{fig:log_term_coef} illustrates the fits to the data for these expressions. The fit parameters are determined with an accuracy of 0.1\% and are given by 
\begin{equation}\label{eq:a_and_b}
	c_2=-\frac{1}{90},\qquad c_3=-\frac{1}{6},\qquad c_4= \frac{1}{6}.
\end{equation}

In conclusion, the results confirm that the UV-divergent part of the entanglement entropy has the form displayed in equation \eqref{eq:genentropy}.

\section{Analytical Results}
\label{sec:analytic}

We next turn to the last term in equation \eqref{eq:genentropy} that arises from the IR sector of the theory. This term was derived for a massless scalar field in a dS background in planar coordinates \cite{Boutivas:2024sat,Boutivas:2024lts}. The size of the system is the only available IR cutoff in that case. In the more general case of a massive field, the mass can also act as a cutoff for the long-range correlations. The IR cutoff will be a combination of the correlation length of the field and the size of the system so that the UV-finite part of the entanglement entropy will have a more complicated structure. As we have already mentioned, the dominance of the UV-divergent terms makes it difficult to extract the structure of the UV-finite terms through the numerical approach. We analyzed our data for the UV-finite part of the entanglement entropy and identified the mass as the effective IR cutoff when the overall system corresponds to the entire spatial sphere. However, extracting the functional dependence of the entropy on the mass has not been possible. 

It is important to note that the role of the mass term is rather different in the current case than in the dS background expressed in planar coordinates. Each constant-time slice in the latter case is non-compact. This means that the field mass acts as an IR cutoff by determining the range over which field fluctuations are substantial. In the current case, the constant-time slice is compact and the role of the mass is to cut off the IR divergence generated by the vanishing energy of the normalizable zero mode in the $\ell=0$ sector, see equation \eqref{eq:zero_mode}. 

In order to bypass the numerical difficulties and also be as close as possible to the setup of \cite{Boutivas:2024sat,Boutivas:2024lts}, we approach the IR issue analytically in the following by considering an overall system that does not cover the entire sphere in the background of the present work. Thus, we are able to obtain a term analogous to the last one in equation \eqref{eq:genentropy}.

\subsection{The Kernels}
\label{subsec:kernels}
As we discussed in section \ref{subsec:Calculation_Entropy},  the entanglement entropy can be calculated using the eigenvalues of the matrix $\mathcal{M}$, defined in equation \eqref{eq:Matrix_M}. Given a coupling matrix $K$, one has to calculate the matrices $\Omega$---the positive square root of matrix $K$---and $\Omega^{-1}$, obtain the kernels that correspond to their continuum limit, and then form the combination $\left(\Omega^{-1}\right)_A\cdot\left(\Omega\right)_A$. Here, the index $A$ indicates that only the corresponding blocks are used, and $\cdot$ stands for integration over the appropriate domain, which is the continuum counterpart of matrix multiplication. We will provide explicit formulae in the following. More details on this approach can be found in \cite{Katsinis:2024gef}.

In order to perform the calculation, we first have to obtain the spectral decomposition of the kernel $K\left(w,w^\prime\right)$, which is the continuous expression corresponding to the coupling matrix $K$. In the continuum limit, the eigenvalue problem for the matrix $K$ assumes the form of \eqref{eq:eigenvalue_prob_cont}, accompanied by Dirichlet boundary conditions. As a result, for $d=3$ and $\ell=0$, we have
\begin{equation}
	K\left(w,w^\prime\right)=\frac{2}{\pi a}\sum_{k=1}^{\infty}\left(\frac{k^2}{a^2}+\tilde{\mu}^2\right)\sin\frac{k w}{a}\sin\frac{k w^\prime}{a},
\end{equation}
where $\tilde{\mu}^2=\mu^2-\frac{1}{a^2}$. Since the massless $(\mu=0)$ theory has a zero mode in its spectrum, we introduce a cutoff $\theta_M$ and define the theory in the interval $[0,a\theta_M]$. Effectively, we impose Dirichlet boundary condition at $\theta=\theta_M$ instead of $\theta=\pi$. In this way, $\mu$ can be set to zero without affecting the consistency of the calculation. The kernel now assumes the form
\begin{equation}
	K\left(w,w^\prime\right)=\frac{2}{a \theta_M}\sum_{k=1}^{\infty}\left(\frac{k^2 \pi^2}{a^2\theta_M^2}+\tilde{\mu}^2\right)\sin\frac{k \pi w}{a\theta_M}\sin\frac{k \pi w^\prime}{a\theta_M}.
\end{equation}
Obviously, at any point of the calculation, we can set $\theta_M$ equal to $\pi$ and recover the original form of the kernel. We point out that the kernel is normalized correctly for a theory in $[0,a\theta_M]$. Moreover, as a consistency check, the kernel can also be written in closed form as
\begin{equation}
	K\left(w,w^\prime\right)=\frac{2}{a \theta_M }\sum_{k=1}^{\infty}\left(-\frac{\partial^2}{\partial w^2}+\tilde{\mu}^2\right)\sin\frac{k \pi w}{a\theta_M}\sin\frac{k \pi w^\prime}{a\theta_M}= \left(-\frac{\partial^2}{\partial w^2}+\tilde{\mu}^2\right)\delta(w-w^\prime),
\end{equation}
where the $\delta$-distribution acts on fields that obey Dirichlet conditions in the aforementioned interval.

Writing the spectral decomposition of $\Omega\left(w,w^\prime\right)$ and $\Omega^{-1}\left(w,w^\prime\right)$ is rather straightforward. We have
\begin{align}
	\Omega\left(w,w^\prime\right)&=\frac{2}{a\theta_M }\sum_{k=1}^{\infty}\left(\frac{k^2 \pi^2}{a^2\theta_M^2}+\tilde{\mu}^2\right)^{1/2}\sin\frac{k \pi w}{a\theta_M}\sin\frac{k \pi w^\prime}{a\theta_M},\\
	\Omega^{-1}\left(w,w^\prime\right)&=\frac{2}{a\theta_M}\sum_{k=1}^{\infty}\left(\frac{k^2 \pi^2}{a^2\theta_M^2}+\tilde{\mu}^2\right)^{-1/2}\sin\frac{k \pi w}{a\theta_M}\sin\frac{k \pi w^\prime}{a\theta_M}.
\end{align}
The kernels can also be written in a form that is more convenient for an expansion around flat space, namely
\begin{align}
	\Omega\left(w,w^\prime\right)&=\frac{2}{\pi}\frac{\pi^2 }{a^2\theta_M^2}\sum_{k=1}^{\infty}\left(k^2-\delta\right)^{1/2}\sin\frac{k \pi w}{a\theta_M}\sin\frac{k \pi w^\prime}{a\theta_M},\\
	\Omega^{-1}\left(w,w^\prime\right)&=\frac{2}{\pi}\sum_{k=1}^{\infty}\left(k^2-\delta\right)^{-1/2}\sin\frac{k \pi w}{a\theta_M}\sin\frac{k \pi w^\prime}{a\theta_M},
\end{align}
where the parameter $\delta$ is defined as
\begin{equation}\label{eq:sqrt_exp}
	\delta = -\frac{a^2\theta_M^2}{\pi^2}\tilde{\mu}^2 =\frac{\theta_M^2}{\pi^2}\left(1-\mu^2a^2\right).
\end{equation}
Notice that $\delta$ is positive between the conformal point $\mu = 1/a$ and the massless limit. Using the expansions
\begin{equation}
	\frac{1}{\sqrt{k^2-\delta}}=\sum _{i=0}^{\infty } \frac{(2 i)!}{\left(2^i i!\right)^2}\frac{\delta^i}{k^{2 i+1}},\qquad \sqrt{k^2-\delta}=-\sum _{i=0}^{\infty } \frac{(2 i)!}{(2 i-1) \left(2^i i!\right)^2}\frac{\delta^i}{k^{2 i-1}},
\end{equation}
which are convergent for $\vert \delta\vert < k^2$ and $\vert \delta\vert \leq k^2$, respectively, we obtain
\begin{align}
	\Omega\left(w,w^\prime\right)&=-\frac{2}{\pi}\frac{\pi^2 }{a^2\theta_M^2}\sum_{k=1}^{\infty}\sum _{i=0}^{\infty } \frac{(2 i)!}{(2 i-1) \left(2^i i!\right)^2}\frac{\delta^i}{k^{2 i-1}}\sin\frac{k \pi w}{a\theta_M}\sin\frac{k \pi w^\prime}{a\theta_M},\\
	\Omega^{-1}\left(w,w^\prime\right)&=\frac{2}{\pi}\sum_{k=1}^{\infty}\sum _{i=0}^{\infty } \frac{(2 i)!}{\left(2^i i!\right)^2}\frac{\delta^i}{k^{2 i+1}}\sin\frac{k \pi w}{a\theta_M}\sin\frac{k \pi w^\prime}{a\theta_M}.
\end{align}
Since we make use of the expansions for any $k\in\mathbb{Z}^{*}$, we must assume that $\vert \delta\vert <1$. 
This allows us to set $\theta_M=\pi$ in the vicinity of the conformal point where $\mu\simeq1/a$. On the other hand, setting $\mu=0$ is possible if the theory is defined only on a portion of the sphere, with $\theta_M$ acting as an IR regulator. In this way the IR regulator can be identified either with the effective mass of the field, or with the size of the overall system.

Interchanging the order of the two summations and identifying the summations over $k$ with polylogarithms, we obtain
\begin{align}
	\Omega\left(w,w^\prime\right)&=-\frac{\pi}{a\theta_M}\sum _{i=0}^{\infty } \frac{(2 i)!}{(2 i-1) \left(2^i i!\right)^2}\delta^i\,\omega_{-2i+1} \left(w,w^\prime\right),\label{eq:Omega_series}\\
	\Omega^{-1}\left(w,w^\prime\right)&=\frac{a\theta_M}{\pi}\sum _{i=0}^{\infty } \frac{(2 i)!}{\left(2^i i!\right)^2}\delta^i\,\omega_{-2i-1} \left(w,w^\prime\right),\label{eq:Omega_Inv_series}
\end{align}
where
\begin{equation}
	\begin{split}\label{eq:omega_def}
		\omega_{-n} \left(w,w^\prime\right)&=\frac{2}{a\theta_M}\sum_{k=1}^\infty\frac{1}{k^n}\sin\frac{k \pi w}{a\theta_M}\sin\frac{k \pi w^\prime}{a\theta_M}\\
		&=\frac{1}{2a\theta_M} \left[\text{Li}_n\left(e^{\frac{i \pi  (w-w^\prime)}{a \theta_M}}\right)-\text{Li}_n\left(e^{\frac{i \pi  (w+w^\prime)}{a \theta_M}}\right)+\mathrm{c.c.}\right].
	\end{split}
\end{equation}
The polylogarithms that appear in these expressions are given by 
\begin{align}
	\mathrm{Li}_{-1}(z)&=\frac{z}{(1-z)^2}\\
	\mathrm{Li}_1(z)&=-\ln(1-z)\\
	\begin{split}
		\mathrm{Li}_n(\exp z)&=\sum _{k=0}^{n-2} \frac{\zeta (n-k)z^k }{k!}-\frac{\left(\ln (-z)-H_{n-1}\right)z^{n-1}}{(n-1)!} \\
		&\hspace*{4.5cm}+\frac{\zeta (0) z^n}{n!}+\sum _{k=1}^{\infty} \frac{\zeta (1-2k) z^{2k+n-1}}{(2k+n-1)!},\quad n\geq2,
	\end{split}
\end{align}
where the last series is convergent for $\vert z\vert<2\pi$. Here, $H_n$ stands for the $n$-th harmonic number and $\zeta$ for Riemann's zeta function. Notice that equation \eqref{eq:omega_def} implies that the kernels $\omega_{-n} \left(w,w^\prime\right)$ are symmetric and that they obey the recursion relation
\begin{equation}\label{eq:derivative_recursion}
	\frac{\partial^2}{\partial w^2}\omega_{-n} \left(w,w^\prime\right)=-\frac{\pi^2}{a^2\theta_M^2}\omega_{2-n} \left(w,w^\prime\right).
\end{equation}

Keeping only the leading and next-to-leading order terms, we obtain
\begin{multline}
	\Omega^{-1}\left(w,y\right)\Omega\left(y,w^\prime\right)= \omega_{-1} \left(w,y\right)\omega_{1} \left(y,w^\prime\right)\\
	+\frac{1}{2}\delta\left[\omega_{-3} \left(w,y\right)\omega_{1} \left(y,w^\prime\right)-\omega_{-1} \left(w,y\right)\omega_{-1} \left(y,w^\prime\right)\right]+\mathcal{O}\left(\delta^2\right),
\end{multline}
which can also be written as
\begin{multline}
	\Omega^{-1}\left(w,y\right)\Omega\left(y,w^\prime\right)= \omega_{-1} \left(w,y\right)\omega_{1} \left(y,w^\prime\right)\\
	-\frac{1}{2}\delta\frac{a^2\theta_M^2}{\pi^2}\frac{\partial}{\partial y}\left[\omega_{-3} \left(w,y\right)\frac{\partial}{\partial y}\omega_{-1} \left(y,w^\prime\right)-\omega_{-1} \left(y,w^\prime\right)\frac{\partial}{\partial y}\omega_{-3} \left(w,y\right)\right]+\mathcal{O}\left(\delta^2\right).
\end{multline}
Thus, the curved-space $\Omega^{-1}\left(w,y\right)\Omega\left(y,w^\prime\right)$ can be written as its flat-space counterpart $\omega_{-1} \left(w,y\right)\omega_{1} \left(y,w^\prime\right)$ plus corrections that can be expressed as a total derivative. This behavior persists to all orders in $\delta$; see Appendix \ref{app:All-order}.

\subsection{Calculation of the Entropy}
The calculation of entanglement entropy uses the eigenvalues of the matrix $\mathcal{M}$, defined in equation \eqref{eq:Matrix_M}. In the continuum limit, this matrix becomes the kernel
\begin{equation}\label{eq:Kernel_M}
	\mathcal{M}\left(w,w^\prime\right)=\int_{0}^{a\theta_R}dy\,\Omega^{-1}\left(w,y\right)\Omega\left(y,w^\prime\right),
\end{equation}
where we assumed that the subsystem $A$ consists of the degrees of freedom in the interval $[0,a\theta_R)$. Besides $y$, both $w$ and $w^\prime$ take values in this interval. It is advantageous to extend the range of integration to $a\theta_M$ in order to isolate a $\delta$-function contribution.\footnote{By definition the kernels $\Omega$ and $\Omega^{-1}$ obey
\begin{equation*}
		\int_{0}^{a\theta_M}dy\,\Omega^{-1}\left(w,y\right)\Omega\left(y,w^\prime\right) = \delta\left(w-w^\prime\right).
\end{equation*}} For this reason, we write  
\begin{equation}
	\mathcal{M}(w,w^\prime)=\delta(w-w^\prime) -\tilde{\mathcal{M}}\left(w,w^\prime\right),
\end{equation}
where
\begin{equation}
	\tilde{\mathcal{M}}\left(w,w^\prime\right)=\int_{a\theta_R}^{a\theta_M}dy\,\Omega^{-1}\left(w,y\right)\Omega\left(y,w^\prime\right).
\end{equation}
Obviously $\mathcal{M}$ and $\tilde{\mathcal{M}}$ share the same eigenfunctions and their eigenvalues are related via $\lambda=1-\tilde{\lambda}$. Thus, equation \eqref{eq:SEE_of_M} assumes the form
\begin{equation}\label{eq:SEE_of_M_tilde}
	S_{\textrm{EE}}=\sum_{i}\left(\frac{\sqrt{1-\tilde{\lambda}_i}+1}{2}\ln\frac{\sqrt{1-\tilde{\lambda}_i}+1}{2}-\frac{\sqrt{1-\tilde{\lambda}_i}-1}{2}\ln\frac{\sqrt{1-\tilde{\lambda}_i}-1}{2}\right),
\end{equation}
where $\tilde{\lambda}_i$ are the eigenvalues of $\tilde{\mathcal{M}}$. In this equation, we dropped the subscript $\ell\vec{m}$ since this section concerns only the vanishing angular momentum sector of the $(3+1)$-dimensional theory. 
Notice that, unless we impose boundary conditions on the eigenfunctions, the spectrum is continuous and the summation must be replaced by integration.

We can now proceed to the main part of the calculation. Within the framework of section \ref{subsec:kernels} we introduce a perturbative expansion for $\tilde{\mathcal{M}}$ in powers of $\delta$ that reads
\begin{equation}
	\tilde{\mathcal{M}}\left(w,w^\prime\right)=\sum_{i=0}^\infty \delta^{i}\,\tilde{\mathcal{M}}^{(i)}\left(w,w^\prime\right).
\end{equation}
We have
\begin{equation}
	\tilde{\mathcal{M}}^{(0)}\left(w,w^\prime\right)= \int_{a\theta_R}^{a\theta_M}dy\, \omega_{-1} \left(w,y\right)\omega_{1} \left(y,w^\prime\right)
\end{equation}
and
\begin{multline}\label{eq:M_1_tilde}
	\tilde{\mathcal{M}}^{(1)}\left(w,w^\prime\right)=-\frac{a^2\theta_M^2}{2\pi^2}\left[\omega_{-1} \left(y,w^\prime\right)\frac{\partial}{\partial y}\omega_{-3} \left(w,y\right)\right.\\
\left.\left.-\omega_{-3} \left(w,y\right)\frac{\partial}{\partial y}\omega_{-1} \left(y,w^\prime\right)\right]\right\vert_{y\rightarrow a\theta_R}.
\end{multline}
On the same token, the entanglement entropy, given by equation \eqref{eq:SEE_of_M_tilde}, is expanded as
\begin{equation}\label{eq:SEE_expansion}
	S_{\textrm{EE}}=\sum_{i=0}^\infty \delta^{i}\,S^{(i)}_{\textrm{EE}}.
\end{equation}
It is straightforward to show that
\begin{equation}
	S^{(0)}_{\textrm{EE}} =\sum_{i}\left(\frac{\sqrt{1-\tilde{\lambda}^{(0)}_i}+1}{2}\ln\frac{\sqrt{1-\tilde{\lambda}^{(0)}_i}+1}{2}-\frac{\sqrt{1-\tilde{\lambda}^{(0)}_i}-1}{2}\ln\frac{\sqrt{1-\tilde{\lambda}^{(0)}_i}-1}{2}\right)
\end{equation}
and
\begin{equation}\label{eq:see_nlo_1}
	S^{(1)}_{\textrm{EE}}=-\sum_{i}\frac{\lambda^{(1)}_i}{2\sqrt{1-\tilde{\lambda}^{(0)}_i}}\arccoth\sqrt{1-\tilde{\lambda}^{(0)}_i},
\end{equation}
where $\tilde{\lambda}^{(0)}_i$ are the eigenvalues of $\tilde{\mathcal{M}}$ at zeroth order in $\delta$, i.e. the eigenvalues of $\tilde{\mathcal{M}}^{(0)}$, and $\tilde{\lambda}^{(1)}_i \delta$ are their leading order corrections. Obviously, $S^{(0)}_{\textrm{EE}}$ is the entanglement entropy of a massless theory in flat space, whereas $S^{(1)}_{\textrm{EE}}\delta$ is the leading correction due to the curvature of the background.

The flat-space eigenvalue problem was solved in \cite{Katsinis:2024gef}; see also \cite{Callan:1994py}. Adapting the notation to match the one of our calculation, the right eigenfunctions $f(x;\omega)$, the left eigenfunctions $g(x;\omega)$, and the eigenvalues $\tilde{\lambda}^{(0)}(\omega)$  of $\tilde{\mathcal{M}}^{(0)}\left(w,w^\prime\right)$ read
\begin{align}
	f(w;\omega)&=\sin\left(\omega u(w)\right),\\
	g(w;\omega)&=\frac{\pi}{2a\theta_M}\frac{\cosh u(w)+\cos\frac{\pi \theta_R}{\theta_M}}{\sin\frac{\pi \theta_R}{\theta_M}}\sin\left(\omega u(w)\right),\\
	\tilde{\lambda}^{(0)}(\omega)&= -\frac{1}{\sinh^2\left(\pi \omega\right)},
\end{align}
where
\begin{equation}
	u(w)=\ln\frac{\sin\frac{\pi\left(a \theta_R+w\right)}{2 a \theta_M}}{\sin\frac{\pi\left(a \theta_R-w\right)}{2 a \theta_M}},\qquad w(u)=\frac{2 a \theta_M}{\pi}\arctan\left(\tanh\frac{u}{2}\tan\frac{\pi \theta_R}{2\theta_M}\right).
\end{equation}
The eigenfunctions are normalized according to
\begin{equation}
	\int_{0}^{a\theta_R}dw\, f(w;\omega)g(w;\omega^\prime)=\frac{\pi}{4}\delta\left(\omega-\omega^\prime\right).
\end{equation}

Making further analytical progress is possible if we assume that the subsystem $A$ is much smaller than the total system, 
so that $\frac{\pi \theta_R}{2\theta_M}\ll 1$. In this limit, we have
\begin{align}
	f(w;\omega)&=\sin\left(\omega u(w)\right),\\
	g(w;\omega)&=\frac{1}{a \theta_R}\cosh^2 \frac{u(w)}{2}\sin\left(\omega u(w)\right),\\
	\tilde{\lambda}^{(0)}(\omega)&= -\frac{1}{\sinh^2\left(\pi \omega\right)},
\end{align}
where
\begin{equation}\label{eq:change_of_variables}
	u(w)=\ln\frac{a \theta_R+w}{a \theta_R-w},\qquad w(u)= a \theta_R\tanh\frac{u}{2}.
\end{equation}
The kernels $\omega_{1}$, $\omega_{-1}$, and $\omega_{-3}$ assume the form
\begin{align}
	\omega_{1}(w,w^\prime)&=\frac{a \theta_M}{\pi^2}\left[\frac{1}{\left(w-w^\prime\right)^2}-\frac{1}{\left(w+w^\prime\right)^2}\right],\\
	\omega_{-1}(w,w^\prime)&=\frac{1}{a \theta_M}\ln\frac{w+w^\prime}{\left\vert w-w^\prime\right\vert},\\
	\begin{split}
		\omega_{-3}(w,w^\prime)&=\frac{1}{a \theta_M}\frac{\pi^2 \theta_R^2}{\theta_M^2}\left[\left(3-2\ln\frac{\pi\theta_R}{\theta_M}\right)\frac{w w^\prime}{a^2\theta_R^2}+\frac{\left(w-w^\prime\right)^2}{2 a^2\theta_R^2}\ln\frac{\left\vert w-w^\prime\right\vert}{ a\theta_R}\right.\\ 
		&\left.\hspace{7cm}-\frac{\left(w+w^\prime\right)^2}{2 a^2\theta_R^2}\ln\frac{\left( w+w^\prime\right)}{ a\theta_R}\right].\label{eq:om_m3_L_inf}
	\end{split}
\end{align}

The leading order corrections to the eigenvalues, $\tilde{\lambda}^{(1)}(\omega)$, are given by
\begin{equation}\label{eq:lambda_2}
	\tilde{\lambda}^{(1)}(\omega)=\frac{\int_{0}^{a \theta_R}dw \int_{0}^{a \theta_R} dw^\prime\tilde{\mathcal{M}}^{(1)}\left(w,w^\prime\right)f(w^\prime;\omega)g(w;\omega)}{\int_{0}^{a \theta_R}dw\,f(w;\omega)g(w;\omega)}.
\end{equation}
The denominator is divergent, and we have to regularize the integral. This is why this factor could not be absorbed by redefining the eigenfunctions $f$ and $g$. In particular, using the change of variable \eqref{eq:change_of_variables} we obtain
\begin{equation}
	\int_{0}^{a \theta_R}dw\,f(w;\omega)g(w;\omega)=\frac{1}{2}\int_{0}^{\infty} du \sin^2\left(\omega u\right).
\end{equation}
One natural way to regularize the integral is to restrict the integration over $w$ to $a \theta_R-\epsilon$, where $\epsilon$ is a UV regulator. Thus, we do not integrate all the way up to the entangling surface, but we stop infinitesimally before it. This regularization scheme is appropriate for dealing with the strong entanglement between adjacent degrees of freedom separated by the entangling surface. With respect to the coordinate $u$, this procedure is equivalent to introducing an upper limit of integration $u_{\textrm{max}}$. In this way, we obtain
\begin{equation}
	\int_{0}^{a \theta_R-\epsilon}dw\,f(w;\omega)g(w;\omega)=\frac{1}{4}u_{\textrm{max}}.
\end{equation}
Equation \eqref{eq:change_of_variables} implies that the two regulators $\epsilon$ and $u_{\textrm{max}}$ are related by
\begin{equation}
	u_{\textrm{max}}=\ln\frac{2 a \theta_R}{\epsilon}.
\end{equation}
Moreover, the UV scale can be utilized to discretize the spectrum according to
\begin{equation}
	\omega_k=\frac{k\pi}{u_{\textrm{max}}},\qquad k\in\mathbb{N}^{*}.
\end{equation}

Making use of equation \eqref{eq:M_1_tilde}, the numerator of equation \eqref{eq:lambda_2} consists of 4 integrals
\begin{align}
	I_1(\omega)=&\int_0^{a\theta_R-\epsilon}dw^\prime\, \omega_{-1} \left(a\theta_R,w^\prime\right) f(w^\prime;\omega),\\
	I_2(\omega)=&\int_0^{a\theta_R-\epsilon}dw^\prime\, \left.\frac{\partial}{\partial y}\omega_{-1} \left(y,w^\prime\right)\right\vert_{y\rightarrow a\theta_R}f(w^\prime;\omega),\\
	I_3(\omega)=&\int_0^{a\theta_R-\epsilon}dw\, \omega_{-3} \left(w,a\theta_R\right) g(w;\omega),\\
	I_4(\omega)=&\int_0^{a\theta_R-\epsilon}dw\, \left.\frac{\partial}{\partial y}\omega_{-3} \left(w,y\right)\right\vert_{y\rightarrow a\theta_R}g(w;\omega),
\end{align}
where we used the same regularization scheme. Using the change of variables \eqref{eq:change_of_variables}, it is evident that the integral $I_1$ is convergent, and it assumes the form
\begin{equation}\label{eq:integral_1}
	\begin{split}
		\lim_{\epsilon\rightarrow0}I_1(\omega)&=\frac{\theta_R}{2\theta_M}\lim_{u_{\textrm{max}}\rightarrow\infty}\int_0^{u_{\textrm{max}}} du^\prime \frac{ u^\prime \sin (\omega u^\prime)}{\cosh^2 \frac{u^\prime}{2}} \\
		&=\pi\frac{\theta_R}{\theta_M}\coth(\pi\omega)\left(\frac{\pi \omega}{\sinh(\pi\omega)}-\frac{1}{\cosh(\pi\omega)}\right),
	\end{split}
\end{equation}
where we used the Fourier transform \eqref{eq:FFT_1} in the calculation.\footnote{The Fourier transformations are typically calculated as principal values. Therefore, the regulator $u_{\textrm{max}}$ fits nicely in this derivation.} Similarly, the integral $I_2$ becomes
\begin{equation}\label{eq:integral_2}
	\begin{split}
		\lim_{\epsilon\rightarrow0}I_2(\omega)&=-\frac{1}{a \theta_M}\lim_{u_{\textrm{max}}\rightarrow\infty}\int_0^{u_{\textrm{max}}} du^\prime \tanh\frac{u^\prime}{2} \sin (\omega u^\prime)\\
		&=-\frac{1}{a \theta_M}\frac{\pi}{\sinh (\pi  \omega )},
	\end{split}
\end{equation}
where we used the Fourier transform \eqref{eq:FFT_2}. Regarding the integrals $I_3$ and $I_4$, we have
\begin{align}
	I_3(\omega)=&\frac{1}{2}\int_0^{u_{\textrm{max}}}du\, \omega_{-3} \left(w(u),a\theta_R\right) \sin(\omega u), \label{eq:integral_3_def}\\
	I_4(\omega)=&\frac{1}{2}\int_0^{u_{\textrm{max}}}du\, \left.\frac{\partial}{\partial y}\omega_{-3} \left(w(u),y\right)\right\vert_{y\rightarrow a\theta_R}\sin(\omega u). \label{eq:integral_4_def}
\end{align}
Putting everything together, equation \eqref{eq:see_nlo_1} assumes the form
\begin{equation}
	S^{(1)}_{\textrm{EE}}=\frac{a^2\theta_M^2}{\pi^2}\frac{\pi}{u_{\textrm{max}}}\sum_{i}\frac{\omega_i}{\coth\left(\pi \omega_i\right)}\left[I_1(\omega_i) I_4(\omega_i)-I_3(\omega_i)I_2(\omega_i)\right].
\end{equation}
At the limit $u_{\textrm{max}}\rightarrow\infty$, the eigenvalues $\omega_i$ become dense and the sum $\frac{\pi}{u_{\textrm{max}}}\sum_{i}$ turns to an integral $\int_0^\infty d\omega$, see \cite{Callan:1994py}. Thus, $S^{(1)}_{\textrm{EE}}$ is given by
\begin{equation}
	S^{(1)}_{\textrm{EE}}=\frac{a^2\theta_M^2}{\pi^2}\int_0^\infty d\omega\frac{\omega}{\coth\left(\pi \omega\right)}\left[I_1(\omega) I_4(\omega)-I_3(\omega)I_2(\omega)\right].
\end{equation}
Substituting $I_1$ and $I_2$ using equations \eqref{eq:integral_1} and \eqref{eq:integral_2}, we have
\begin{equation}
	S^{(1)}_{\textrm{EE}}=\frac{a \theta_M}{\pi}\int_0^\infty d\omega\,\omega\left[\frac{1}{\cosh\left(\pi \omega\right)}I_3(\omega)+a\theta_R\left(\frac{\pi \omega}{\sinh\left(\pi \omega\right)}-\frac{1}{\cosh\left(\pi \omega\right)}\right)I_4(\omega)\right].
\end{equation}
Then, equations \eqref{eq:integral_3_def} and \eqref{eq:integral_4_def} imply that $S^{(1)}_{\textrm{EE}}$ is given by
\begin{multline}
	S^{(1)}_{\textrm{EE}}=\frac{a \theta_M}{2\pi}\int_0^{\infty}du\int_0^\infty d\omega\frac{\omega\sin(\omega u)}{\cosh\left(\pi \omega\right)}\left[\omega_{-3} \left(w(u),a\theta_R\right)\phantom{\left(\frac{\pi \omega}{\tanh\left(\pi \omega\right)}-1\right)}\right. \\ \left.+a\theta_R\left(\frac{\pi \omega}{\tanh\left(\pi \omega\right)}-1\right)\left.\frac{\partial}{\partial y}\omega_{-3} \left(w(u),y\right)\right\vert_{y\rightarrow a\theta_R}\right],
\end{multline}
where we interchanged the order of the integrations. The integrals over $\omega$ can be performed using equations \eqref{eq:FFT_1} and \eqref{eq:FFT_3}, resulting in
\begin{multline}\label{eq:S1_integral}
	S^{(1)}_{\textrm{EE}}=-\frac{a \theta_M}{4\pi}\int_0^{\infty}du\left[\omega_{-3} \left(w(u),a\theta_R\right)\frac{\partial}{\partial u} \frac{1}{\cosh\frac{u}{2}} \right. \\ \left.
	+a \theta_R\left.\frac{\partial}{\partial y}\omega_{-3} \left(w(u),y\right)\right\vert_{y\rightarrow a\theta_R}\frac{\partial}{\partial u}\frac{\frac{\pi}{2} -\cosh \frac{u}{2} }{\cosh^2\frac{u}{2}} \right].
\end{multline}
We integrate by parts. There is no boundary term\footnote{Given that $w(0)=0$, we have
\begin{equation}
	\omega_{-3} \left(0,a\theta_R\right)=\left.\frac{\partial}{\partial y}\omega_{-3} \left(0,y\right)\right\vert_{y\rightarrow a\theta_R}=0.
\end{equation}
Moreover, since $\lim_{u\rightarrow\infty}w(u)=a\theta_R$, we also have
\begin{align}
	\lim_{u\rightarrow\infty}\omega_{-3} \left(w(u),a\theta_R\right)&=\frac{1}{a \theta_M}\frac{\pi^2 \theta_R^2}{\theta_M^2}\left(3-2\ln\frac{2\pi\theta_R}{\theta_M}\right)\\
	\lim_{u\rightarrow\infty}\left.\frac{\partial}{\partial y}\omega_{-3} \left(w(u),y\right)\right\vert_{y\rightarrow a\theta_R}&=\frac{1}{a \theta_M}\frac{\pi^2}{a^2\theta_M^2}\frac{2\theta_R}{\theta_R}\left(1-\ln\frac{2\pi\theta_R}{\theta_M}\right).
\end{align}} and we obtain
\begin{equation}\label{eq:S1_sum}
	S^{(1)}_{\textrm{EE}}=I_A + I_B,
\end{equation}
where the integrals $I_A$ and $I_B$ are given by
\begin{align}
	I_A&=\frac{a \theta_M}{4\pi}\int_0^{\infty}du\frac{1}{\cosh \frac{u}{2}}\frac{\partial}{\partial u}\omega_{-3} \left(w(u),a\theta_R\right), \label{eq:S1_integral_A}\\
	I_B&=\frac{a^2 \theta_M^2}{4\pi}\frac{\theta_R}{\theta_M} \int_0^{\infty}du \frac{\frac{\pi}{2} -\cosh \frac{u}{2} }{\cosh^2\frac{u}{2}} \left.\frac{\partial^2}{\partial y\partial u}\omega_{-3} \left(w(u),y\right)\right\vert_{y\rightarrow a\theta_R}.\label{eq:S1_integral_B}
\end{align}
Substituting equation \eqref{eq:om_m3_L_inf}, we obtain
\begin{align}
	I_A&=\frac{\pi}{4}\frac{\theta_R^2}{\theta_M^2}\int_0^{\infty}du\frac{1}{\cosh^3 \frac{u}{2}} \left(\ln \frac{\theta_M}{2 \pi  \theta_R}+\ln \left(1+e^{-u}\right)+1+\frac{u}{e^u+1}\right), \label{eq:S1_integral_A_2}\\
	I_B&=\frac{\pi}{4}\frac{\theta_R^2}{\theta_M^2} \int_0^{\infty}du \frac{\frac{\pi}{2} -\cosh\frac{u}{2} }{\cosh^4\frac{u}{2}}  \left(\ln \frac{\theta_M}{2 \pi \theta_R}+\ln \left(1+e^{-u}\right)+\frac{u}{2}\right).\label{eq:S1_integral_B_2}
\end{align}
Performing the integrations, we find 
\begin{equation}
	S^{(1)}_{\textrm{EE}}=\frac{1}{6}\frac{\pi^2\theta_R^2}{\theta_M^2}\left(\ln \frac{\theta_M}{2 \pi \theta_R}+\frac{4}{3}\right).
\end{equation}

In total, the entanglement entropy,vdefined in equation \eqref{eq:SEE_expansion}, becomes
\begin{equation} \label{eq:finalresult}
	S_{\textrm{EE}}=S^{(0)}_{\textrm{EE}}+\frac{1}{6}\theta_R^2\left(1-\mu^2a^2\right)\left(\ln \frac{\theta_M}{2 \pi \theta_R}+\frac{4}{3}\right)+\mathcal{O}\left(\delta^2\right)+\mathcal{O}\left(\delta \pi^3 \frac{\theta_R^3}{\theta_M^3}\right).
\end{equation}
The leading term corresponds to the flat-space entanglement entropy of a massless scalar field in $1+1$ dimensions, given by the well-known formula \cite{Holzhey:1994we,Calabrese:2004eu}
\begin{equation}
S^{(0)}_{\textrm{EE}}=\frac{1}{6}\ln\frac{2a\theta_R}{\epsilon}.
\end{equation}
For the $(3+1)$-dimensional theory, the additional UV-divergent contributions from the sectors with $\ell>0$ result in the UV structure of equation \eqref{eq:genentropy}. Also, it has been found numerically in the dS case that the analog of the term $\sim \theta_R^2 \log \theta_R$ has a very small---probably vanishing---coefficient in the full $(3+1)$-dimensional theory \cite{Boutivas:2024lts}. However, the term  $\sim \theta_R^2 \log \theta_M$ exists in the full theory, with the coefficient given by the analytical treatment of the $\ell=0$ sector \cite{Boutivas:2024sat,Boutivas:2024lts}. For the dS background this coefficient is $c_{\rm IR}=\frac{1}{3}$, to be compared with the value  $c_{\rm IR}=\frac{1}{6}$ we obtained for the $\mathbb{R}\times$S$^3$ geometry.
 
We emphasize that, despite the absence of an obvious obstruction, one cannot set $\mu=0$ and $\theta_M=\pi$ at the same time in equation \eqref{eq:finalresult}. This result was obtained under the assumption $\delta=\frac{\theta_M^2}{\pi^2}\left(1-\mu^2a^2\right)<1$, which excludes this parameter choice. The structure of the entanglement entropy when the overall system is the entire sphere and the field mass approaches zero is more complicated than our result, which applies either to a theory on the entire sphere near the conformal point $\mu a \sim 1$, or to the massless theory on a portion of the sphere. 

\section{Discussion}
\label{sec:discussion}

In previous works, we applied our method for calculating entanglement entropy to a scalar theory in several interesting backgrounds, including Minkowski, dS, and AdS space. In the current work, we considered another case, that of the Einstein universe with an $\mathbb{R}\times$S$^3$ geometry. Apart from establishing the validity of our approach for yet another space, we had certain specific points in mind that arose in our previous studies. 
 
The first point concerns the form of the leading UV-divergent term and its dependence on the regularization implemented through the discretization of the theory. The metric \eqref{eq:metricsphere} that we considered differs from the AdS metric in global coordinates by an overall conformal factor. The radial variable $w$, appearing in both metrics, is discretized in both cases according to \eqref{eq:discretization_1}. The leading UV-divergent term was found with high precision to have exactly the same form $\sim \sin^2 \frac{w_R}{a}$ in both cases. In the current case, this reproduces correctly the proper area of the entangling surface, and the area law is satisfied as in flat space. Moreover, a similar factor appears in the UV-divergent logarithmic term, in accordance with equation \eqref{eq:genentropy}. In the AdS case, there seems to be a discrepancy in the leading divergence, as the proper area scales $\sim \tan^2 \frac{w_R}{a}$. However, the correct factor, proportional to the proper area, appears in the logarithmic divergence \cite{Boutivas:2025ksp}. 
 
The comparison of the two cases confirms the conclusions reached in \cite{Boutivas:2025ksp}. The coefficient of the logarithmic term is independent of the particular regulator. This is well established in flat space but also in curved backgrounds, where additional terms proportional to the proper area appear. The leading UV divergence scales with the inverse square of the cutoff, but the coefficient may vary depending on the regularization. In extreme cases, such as the discretized version of the scalar theory in AdS in terms of the global coordinate $w$, the dependence on the area of the entangling surface may be masked. Accounting carefully for the density of degrees of freedom in the radial direction restores the correct dependence, as discussed in the introduction. 
 
For spherical configurations in the spaces we considered, the logarithmic UV divergence can be parameterized as in equation \eqref{eq:genentropy}. The coefficients have been determined numerically to be the following rational numbers with an accuracy of 0.1\%:
 \begin{equation}\label{eq:eu}
 	c_2=-\frac{1}{90},\qquad c_3=-\frac{1}{6},\qquad c_4= \frac{1}{6}
 \end{equation} 
for the Einstein universe,
 \begin{equation}\label{eq:ds}
	c_2=-\frac{1}{90},\qquad c_3=-\frac{1}{6},\qquad c_4= \frac{1}{3}
\end{equation} 
for a dS background \cite{Boutivas:2024lts} \footnote{The value of $c_3$ is a conjecture, based on the requirement that the coefficient of the logarithmic term match that on a flat background for the conformal theory. It is also expected that the value of $c_3$ is independent of the curvature of the background.}, and
 \begin{equation}\label{eq:ads}
	c_2=-\frac{1}{90},\qquad c_3=-\frac{1}{6},\qquad c_4= -\frac{1}{3}
\end{equation} 
for an AdS background \cite{Boutivas:2025ksp}. In all three cases, the values of $c_3$ and $c_4$ are such that the corresponding contributions cancel for a field whose mass arises from a non-minimal coupling to gravity through a term $\frac{1}{6} {\cal R} \phi^2$. The coefficient of the logarithmic term for the resulting Weyl-invariant theory is given by $c_2$ and is related to the $A$-type conformal anomaly \cite{Solodukhin:2008dh}.

The second point we considered in this work is the effect on the entanglement entropy of long-range correlations present in theories where IR divergences may appear. The motivation was the derivation in \cite{Boutivas:2024sat} of the last term in equation \eqref{eq:genentropy} for a massless scalar field in dS space in planar coordinates in the Bunch-Davies vacuum. In the $\mathbb{R}\times$S$^3$ geometry that we considered, IR divergences appear because of the presence of a normalizable zero mode in the spectrum of the scalar field. We performed an explicit analytical calculation along the lines of \cite{Boutivas:2024sat}. The final result, given by equation \eqref{eq:finalresult}, confirms that such a term arises from the $\ell=0$ sector of the theory that is dominated by the zero mode. If an IR cutoff is implemented by restricting the overall theory to a portion of the spatial sphere so that the field mass can be set to zero, the structure of the IR term will be given by equation \eqref{eq:genentropy}, similarly to the dS space. The only difference lies in the value of the coefficient,  which is  $c_{\rm IR}=\frac{1}{6}$, as compared to $c_{\rm IR}=\frac{1}{3}$ in dS.

The combination of the results summarized above generates a consistent picture in which the UV and IR contributions seem to be well understood. This provides the necessary framework in order to shift the focus towards the finite part of the entropy, which is related to fundamental issues, such as the relation between the entanglement entropy and the $a$-theorem \cite{Casini:2017vbe,Abate:2024nyh}. One obstacle in this direction is the extreme accuracy required for a numerical calculation that could isolate the finite part of the entropy. The current work has used 33--35 significant digits of precision for the numerical calculation on a radial lattice with up to 450 points. Much larger lattices will be required for the study of the finite part. Moreover, some intuition is necessary on the functional form of entropy and its dependence on parameters such as the field mass and the length scale of the background. Despite these difficulties, a calculation seems feasible and will be the subject of future work. 

\appendix
\section{All-order Results}
\label{app:All-order}
	In this appendix we calculate $\Omega^{-1}\left(w,y\right)\Omega\left(y,w^\prime\right)$ to all orders in $\delta$. The final formula is suitable for integration over $y$.
	
	Implementing the Cauchy product of the series appearing in equations \eqref{eq:Omega_series} and \eqref{eq:Omega_Inv_series} we have
	\begin{equation}
		\begin{split}
			\Omega^{-1}\left(w,y\right)\Omega\left(y,w^\prime\right)&=-\sum _{i=0}^{\infty } \sum _{j=0}^{\infty } \frac{(2 i)!(2 j)!\delta^{i+j}}{\left(2^i i!\right)^2(2 j-1) \left(2^j j!\right)^2}\omega_{-2i-1} \left(w,y\right)\omega_{-2j+1} \left(y,w^\prime\right)\\
			&=-\sum _{j=0}^{\infty }\delta^{j}\sum _{i=0}^{j} \frac{(2 i)!(2 (j-i))!}{\left(2^i i!\right)^2(2 (j-i)-1) \left(2^{(j-i)} (j-i)!\right)^2} \\
			&\hspace*{4.75cm}\times\omega_{-2i-1} \left(w,y\right)\omega_{-2(j-i)+1} \left(y,w^\prime\right).
		\end{split}
	\end{equation}
Even though this expression looks cumbersome, a significant simplification can be achieved because, for $j\geq 1$, the sum over $i$ obeys the identity
	\begin{multline}
		\sum _{i=0}^{j} \frac{(2 i)!(2 (j-i))!}{\left(2^i i!\right)^2(2 (j-i)-1) \left(2^{(j-i)} (j-i)!\right)^2} \omega_{-2i-1} \left(w,y\right)\omega_{-2(j-i)+1} \left(y,w^\prime\right)\\
		=\frac{1}{j\pi}\sum _{i=0}^{j-1} \frac{\Gamma\left(i+\frac{3}{2}\right)}{i!} \frac{\Gamma\left(j-i-\frac{1}{2}\right)}{(j-1-i)!}
		\left[ \omega_{-2i-1} \left(w,y\right)\omega_{-2(j-i)+1} \left(y,w^\prime\right)\right. \\
		\left.- \omega_{-2i-3} \left(w,y\right)\omega_{-2(j-i)+3} \left(y,w^\prime\right)\right].
	\end{multline}
Then, equation \eqref{eq:derivative_recursion} implies that the square bracket is a total derivative with respect to $y$. Indeed, we have
	\begin{multline}
		\omega_{-2i-1} \left(w,y\right)\omega_{-2(j-i)+1} \left(y,w^\prime\right)- \omega_{-2i-3} \left(w,y\right)\omega_{-2(j-i)+3} \left(y,w^\prime\right) =-\frac{a^2\theta_M^2}{\pi^2}\\
		\times\left[\omega_{-2(j-i)+1} \left(y,w^\prime\right)\frac{\partial^2}{\partial y^2}\omega_{-2i-3} \left(w,y\right)- \omega_{-2i-3} \left(w,y\right)\frac{\partial^2}{\partial y^2}\omega_{-2(j-i)+1} \left(y,w^\prime\right)\right]
	\end{multline}
	or
	\begin{multline}
		\omega_{-2i-1} \left(w,y\right)\omega_{-2(j-i)+1} \left(y,w^\prime\right)- \omega_{-2i-3} \left(w,y\right)\omega_{-2(j-i)+3} \left(y,w^\prime\right) =-\frac{a^2\theta_M^2}{\pi^2}\\
		\times\frac{\partial}{\partial y}\left[\omega_{-2(j-i)+1} \left(y,w^\prime\right)\frac{\partial}{\partial y}\omega_{-2i-3} \left(w,y\right)- \omega_{-2i-3} \left(w,y\right)\frac{\partial}{\partial y}\omega_{-2(j-i)+1} \left(y,w^\prime\right)\right].
	\end{multline}
	Putting everything together, $\Omega^{-1}\left(w,y\right)\Omega\left(y,w^\prime\right)$ is given by
	\begin{multline}
		\Omega^{-1}\left(w,y\right)\Omega\left(y,w^\prime\right)=\omega_{-1} \left(w,y\right)\omega_{1} \left(y,w^\prime\right)\\+\frac{a^2\theta_M^2}{\pi^2}\frac{\partial}{\partial y}\sum _{j=1}^{\infty }\frac{\delta^{j}}{j\pi}\sum _{i=0}^{j-1} \frac{\Gamma\left(i+\frac{3}{2}\right)}{i!} \frac{\Gamma\left(j-i-\frac{1}{2}\right)}{(j-1-i)!}\left[\omega_{-2(j-i)+1} \left(y,w^\prime\right)\frac{\partial}{\partial y}\omega_{-2i-3} \left(w,y\right)\right. \\
		\left.- \omega_{-2i-3} \left(w,y\right)\frac{\partial}{\partial y}\omega_{-2(j-i)+1} \left(y,w^\prime\right)\right].
	\end{multline}

\section{Fourier Transforms}
This section lists a family of Fourier transforms used in the derivation of section \ref{sec:analytic}. They can be found in the standard reference \cite{gradshteyn2007table}.
For $\vert a\vert < \vert b\vert$, we have
\begin{align}
	&\int_{-\infty}^{\infty} d\omega \frac{\sinh(a\omega)}{\sinh(b\omega)} e^{i\omega u^\prime}=\frac{\pi}{b}\frac{\sin\frac{\pi a}{b}}{\cos\frac{\pi a}{b}+\cosh\frac{\pi u^\prime}{b}}, \label{eq:FFT_1}\\
	&\int_{-\infty}^{\infty} d\omega \frac{\cosh(a\omega)}{\sinh(b\omega)} e^{i\omega u^\prime}=i\frac{\pi}{b}\frac{\sinh\frac{\pi u^\prime}{b}}{\cos\frac{\pi a}{b}+\cosh\frac{\pi u^\prime}{b}},\label{eq:FFT_2}\\
	&\int_{-\infty}^{\infty} d\omega \frac{\cosh(a\omega)}{\cosh(b\omega)} e^{i\omega u^\prime}=\frac{2\pi}{b}\frac{\cos\frac{\pi a}{2b}\cosh\frac{\pi u^\prime}{2b}}{\cos\frac{\pi a}{b}+\cosh\frac{\pi u^\prime}{b}}, \label{eq:FFT_3}\\
	&\int_{-\infty}^{\infty} d\omega \frac{\sinh(a\omega)}{\cosh(b\omega)} e^{i\omega u^\prime}=i\frac{2\pi}{b}\frac{\sin\frac{\pi a}{2b}\sinh\frac{\pi u^\prime}{2b}}{\cos\frac{\pi a}{b}+\cosh\frac{\pi u^\prime}{b}}, \label{eq:FFT_4}
\end{align}

\bibliographystyle{JHEP} 
\bibliography{sphere_bib}

\end{document}